\documentclass[
aps,
pra,
showpacs
amsmath,
amssymb,
onecolumn,
floatfix,
nobibnotes, 
longbibliography,
lengthcheck,
superscriptaddress,
nofootinbib
]{revtex4-1}

\usepackage{graphicx}
\usepackage{floatflt,epsfig} 
\usepackage{mathtools}
\usepackage{color}
\usepackage{braket}
\usepackage{hyperref}
\usepackage[normalem]{ulem}
\usepackage[english]{babel}
\usepackage{blindtext}
\usepackage{lipsum}  
\usepackage{amsthm}
\usepackage{tabularx}
\usepackage{bm}
\usepackage{dsfont}
\usepackage{bbold}
\usepackage{ulem}
\usepackage[dvipsnames]{xcolor}
\usepackage{soul}

\usepackage[utf8]{inputenc}
\usepackage[english]{babel}

\definecolor{green1}{rgb} {0.658, 0.188, 0.678}

\newcommand {\br} {\mathbf{r}}
\newcommand {\imagi} {\mathrm{i}}
\newcommand {\diff} {\mathrm{d}}

\begin{document}

\title{Relevance of the quadratic diamagnetic and self-polarization terms in cavity quantum electrodynamics}

  \author{Christian Sch\"afer}
  \email[Electronic address:\;]{christian.schaefer@mpsd.mpg.de}
  \affiliation{Max Planck Institute for the Structure and Dynamics of Matter and Center for Free-Electron Laser Science \& Department of Physics, Luruper Chaussee 149, 22761 Hamburg, Germany}
  \author{Michael Ruggenthaler}
  \email[Electronic address:\;]{michael.ruggenthaler@mpsd.mpg.de}
  \affiliation{Max Planck Institute for the Structure and Dynamics of Matter and Center for Free-Electron Laser Science \& Department of Physics, Luruper Chaussee 149, 22761 Hamburg, Germany}
  \author{Vasil Rokaj}
  \email[Electronic address:\;]{vasil.rokaj@mpsd.mpg.de}
  \affiliation{Max Planck Institute for the Structure and Dynamics of Matter and Center for Free-Electron Laser Science \& Department of Physics, Luruper Chaussee 149, 22761 Hamburg, Germany}
  \author{Angel Rubio}
  \email[Electronic address:\;]{angel.rubio@mpsd.mpg.de}
  \affiliation{Max Planck Institute for the Structure and Dynamics of Matter and Center for Free-Electron Laser Science \& Department of Physics, Luruper Chaussee 149, 22761 Hamburg, Germany}
  \affiliation{Nano-Bio Spectroscopy Group,  Dpto. Fisica de Materiales, Universidad del Pa\'is Vasco, 20018 San Sebasti\'an, Spain}
  
\date{\today}

\begin{abstract}
\noindent
Experiments at the interface of quantum-optics and chemistry have revealed that strong coupling between light and matter can substantially modify chemical and physical properties of molecules and solids. While the theoretical description of such situations is usually based on non-relativistic quantum electrodynamics, which contains quadratic light-matter coupling terms, it is commonplace to disregard these terms and restrict
to purely bilinear couplings. 
In this work we clarify the physical origin and the substantial impact of the most common quadratic terms, the diamagnetic and self-polarization terms, and highlight why neglecting them can lead to rather unphysical results. Specifically we demonstrate its relevance by showing that neglecting it leads to the loss of gauge invariance, basis-set dependence, disintegration (loss of bound states) of any system in the basis set-limit, unphysical radiation of the ground state and an artificial dependence on the static dipole. Besides providing important guidance for modeling strongly coupled light-matter systems, the presented results do also indicate under which conditions those effects might become accessible.
\end{abstract}

\date{\today}

\maketitle



Driven by substantial experimental progress in the field of cavity-modified chemistry~\cite{hutchison2012, coles2014b,thomas2016,Zhong2017,Munkhbat2018, Xiang2018, Stranius2018, Wang2019, Peters2019,Ojambati2019,Lather2019}, theoretical methods at the border between quantum-chemical \textit{ab initio} methods and optics have become the focus of many recent investigations~\cite{ruggenthaler2017b,flick2017,schafer2019modification,schafer2018insights,flick2017c,flick2017b,Agranovich2003,Michetti2005,ruggenthaler2011b,GonzlezTudela2013,tokatly2013,buhmann2013dispersionI,buhmann2013dispersionII,ruggenthaler2014,galego2015,del2015quantum,schachenmayer2015,salam2016non,kowalewski2016,herrera2016,Zeb2017,luk2017multiscale,liberato2017virtual,Vendrell2018,FRibeiro2018,hoffmann2018light,hagenmuller2018,fregoni2018manipulating,galego2019cavity,delPino2018,Strathearn2018,Abedi2018,hoffmann2019capturing,Rivera2019,Triana2019,Reitz2019,mordovinaCC,hoffmann2019benchmarking,flick2019light,du2018theory,ribeiro2018polariton,feist2017polaritonic,kockum2019ultrastrong,groenhof2019tracking,campos2019resonant,semenov2019electron,baranov2019ultrastrong}. 
The high complexity of a molecular system, which can undergo, e.g., chemical reactions or quantum phase-transitions, coupled strongly to photons makes the use of some sort of approximation strategy necessary. A common approach is to use approximation strategies designed for atomic two-level-like systems in high-quality optical cavities~\cite{dicke1954,jaynes1963,garraway2011} and to apply them to the quite different situation of molecular systems.
However, under the generalized conditions of cavity-modified chemistry usually disregarded contributions in the theoretical description, e.g., quadratic coupling terms between light and matter, can become important~\cite{rokaj2017,schafer2018insights,viehmann2011superradiant} and might even dominate the physical properties~\cite{flick2017,schafer2018insights,schafer2019modification,flick2017c,sentef2018,hagenmueller2019enhancement}. While the existence of these quadratic terms is well-known~\cite{power1959coulomb,woolley1980gauge,babiker1983derivation,loudon1988,baxter1993canonical,lembessis1993theory,craig1998,spohn2004} their origin, interpretation and consequences are less clear, and when to include them has become the subject of recent intense discussions~\cite{vukics2014,todorov2014dipolar,george2016,rokaj2017,schafer2018insights,bernadrdis2018breakdown,galego2019cavity,DeBernardis2018,andrews2018perspective,DiStefano2019,hernandez2019multi,li2020electromagnetic}. In this work we will elucidate these terms for the most relevant setting of cavity-modified chemistry, i.e., in Coulomb gauge and in the long-wavelength limit, clarify their origins, physical interpretations and consequences as well as show under which conditions and for which observables they become relevant. This will also highlight a domain of applicability of common approximations that disregard these quadratic terms and at the same time indicates under which conditions substantial influence can be expected~\cite{ruggenthaler2017b}, accessible with \textit{ab initio} techniques such as quantum-electrodynamic density-functional theory (QEDFT)~\cite{ruggenthaler2014,ruggenthaler2017b,tokatly2013,flick2015,flick2017c}. Before we do so, let us briefly outline the theory we consider and collect a set of fundamental conditions we deem important for a reasonable theoretical description.

Any theory we employ to model coupled light-matter systems should obey certain fundamental constraints. Which ones these are often depends on the specific situation we consider. For instance, in the case of high-energy physics an adherence to special relativity (physical laws should be Lorentz invariant) is paramount and hence the use of Dirac's equation becomes necessary to capture the behavior of electrons. If we further want to ensure that all interactions among the electrons are local and our theory should stay invariant under local phase transformations we find the Maxwell field coupled to Dirac's momentum operator in a linear (minimal) fashion. However, the resulting theory - which, if quantized, is called quantum electrodynamics (QED) and perfectly describes high-energy scattering events - has many subtle issues~\cite{baez2014introduction}. For low-energy physics a simplified version, where instead of the relativistically invariant momentum the non-relativistic momentum is employed, has been shown to be able to resolve many of these issues~\cite{spohn2004}. The resulting theory of non-relativistic QED (also sometimes called molecular QED~\cite{craig1998,loudon1988,cohen1997photons,salam2010molecular}) is ideally suited to describe atoms, molecules or solids interacting with the quantized light field~\cite{Bach1999,Dereziski2001,HIDAKA2010}. The coupling between light and matter is, however, only defined up to a phase and we need to make a specific choice for this phase, i.e., we need to fix a gauge. 
Changing the gauge or performing a local unitary transformation should not modify physical observables but merely affect their representation in terms of canonical coordinates. While gauge independence is respected by non-relativistic QED, this constraint is specifically challenging for dimensionally-reduced, simplified models~\cite{schafer2018insights,bernadrdis2018breakdown, DiStefano2019,li2020electromagnetic}.
Beside gauge independence, non-relativistic QED guarantees a set of further intuitive and essential conditions. For instance, the physical observables are independent of the chosen coordinate system (or in more quantum-chemical terms, where a specific spatial basis is just one of many basis-set choices, basis-set independent) and it also guarantees the stability of matter, i.e., atoms and molecules are stable if coupled to the vacuum of the electromagnetic field~\cite{spohn2004}. A direct consequence of this fundamental condition is that the combined ground state of light and matter has a zero transversal electric-field expectation value. 
If this would not be the case the system could emit photons and lower its energy. To summarize, a few basic constraints we want a theory of light-matter interactions to adhere to are: All physical observables should be independent of the gauge choice and of the choice of coordinate system (for instance, it would be unphysical that the properties of atoms and molecules would depend on the choice of the origin of the laboratory reference frame), the theory should support stable ground states (else we could not define equilibrium properties and identify specific atoms and molecules) and the coupled light-matter ground state should have a zero transversal electric field (else the system would radiate and cascade into lower-energy states).

In the following we will introduce non-relativistic QED and some of its unitarily equivalent realizations,
highlight the physical implications of the associated transformations and further approximations that lead to the non-relativistic QED in the long-wavelength limit in Sec.~\ref{sec:Maxwell}, and illustrate that the aforementioned fundamental physical conditions will not be retained when disregarding quadratic components in Sec.~\ref{Necessity}. Finally we discuss implications and perspectives in Sec.~\ref{sec:summary}. 
We provide further details in the Appendix. In Appendix~\ref{app:nlimit}, the basic approximations leading to non-relativistic QED are presented, App.~\ref{app:pzw} and \ref{app:trbasis} provide additional details that complement our discussion regarding the Power-Zienau-Wooley transformation and transversal basis functions, and in App.~\ref{app:spectralfeatures} we discuss some implications that go hand in hand with approximating operators.
Our discussions will be presented first in a field-theoretical convention where the four vector potential $A^{\mu}$ is given in Volts and the four charge-current density $j^\mu $ is given in Coulomb per meter squared per second and later in atomic units. By multiplying $A^{\mu}$ by $1/c$ we find the standard convention in terms of Volt second per meter.

\section{From microscopic to macroscopic Maxwell's Equations}
\label{sec:Maxwell}

Classical electrodynamics is at the heart of QED. Consider for instance the inhomogeneous microscopic Maxwell equations \cite{cohen1997photons}
\begin{align*}
\nabla \times \textbf{E}(\textbf{r},t) &= -\partial_t \textbf{B}(\textbf{r},t)\\
\nabla \times \textbf{B}(\textbf{r},t) &= \frac{1}{c^2}\left[\partial_t \textbf{E}(\textbf{r},t)  + \mu_0 c^2 \textbf{j}(\textbf{r},t) \right]~.
\end{align*} 
This representation of light-matter coupling is by no means unique and many different formulations, such as the Riemann-Silberstein~\cite{Gersten-1999,Keller} or the macroscopic Maxwell's equations, have been developed over the years.
To arrive at the latter, let us first rewrite $\mathbf{j}_{} = \mathbf{j}_{}^{\rm b} + \mathbf{j}_{}^{f}$, where $\mathbf{j}_{}^{\rm b}$ is a bound and $\mathbf{j}_{}^{\rm f}$ a free current, and define a bound charge current
\begin{align*}
\mathbf{j}_{}^{\rm b} (\br,t) = \mathbf{\nabla} \times \mathbf{M}(\br,t) + \partial_t \mathbf{P}_{}(\br,t),
\end{align*}
with $\mathbf{M}$ the magnetization and $\mathbf{P}_{}$ the polarization of the matter system. If we then define the displacement field $\mathbf{D} = \epsilon_0 \mathbf{E}_{} + \mathbf{P}_{}$ and the magnetization field $\mathbf{H} = \tfrac{1}{\mu_0} \mathbf{B} - \mathbf{M}$ we end up with
\begin{align*}
-\partial_t \mathbf{D}(\br,t) + \mathbf{\nabla} \times \mathbf{H}(\br,t) = \mathbf{j}^{f}_{}(\br,t) 
\end{align*}
which takes the back-reaction of a given medium on the electromagnetic field into account by the constitutive equations.
Clearly, the classical description of electromagnetic interaction can take different forms for which, without further simplifications, none is superior over the other. These forms deviate merely in their choice of canonical variables, the very same variables that will be quantized to reach QED. The electromagnetic field energy 
\begin{align}
\label{eq:MaxwellEnergy_start}
\mathcal{H}_{\rm{em}}(t) = \frac{\epsilon_{0}}{2} \int \mathrm{d}^3 r
\left( \mathbf{E}^2_{}(\br,t) + c^2 \mathbf{B}^2(\br,t) \right)
\end{align}
is of quadratic form and therefore substituting $\mathbf{D} = \epsilon_0 \mathbf{E}_{} + \mathbf{P}_{}$ into Eq.~\eqref{eq:MaxwellEnergy_start} will naturally lead to quadratic \emph{self-polarization} $\mathbf{P}_{}^2$ and \emph{self-magnetization} $\mathbf{M}^2$ terms.\\

While many equivalent ways of formulating the Maxwell's equations exist, there will be accordingly also several (unitarily equivalent) forms of the resulting non-relativistic QED Hamiltonian. Let us in the following see how this equivalence in QED manifests.
A relativistic quantization procedure with subsequent nonrelativistic limit, as illustrated in App.~\ref{app:nlimit}, is indeed equivalent to introducing the covariant derivative for the electronic system and then quantizing the resulting gauge field~\cite{spohn2004,ruggenthaler2014,ruggenthaler2017b}. 
This minimal-coupling procedure makes the invariance under local phase transformations $\Psi' = e^{\imagi\theta(\textbf{r})}\Psi$ explicit and with the Coulomb gauge condition $\nabla \cdot \hat{\textbf{A}}(\textbf{r})=0$ fixes the local phase $\theta(\textbf{r})$ uniquely. The momentum of each particle is shifted according to $-\imagi \hbar \mathbf{\nabla} \rightarrow (-\imagi \hbar \mathbf{\nabla} - \tfrac{q}{c} \mathbf{\hat{A}}(\br) )$, where $q$ is the charge of the particle and the quantized vector potential is
\begin{align}
\mathbf{\hat{A}}(\mathbf{r})&=\sqrt{\frac{\hbar c^2}{\epsilon_0} }\sum_{\mathbf{n},\lambda}\frac{1}{\sqrt{2\omega_\mathbf{n}}}\left[ \hat{a}_{\mathbf{n},\lambda}\bm{S}_{\textbf{n},\lambda}(\br)+\hat{a}^{\dagger}_{\mathbf{n},\lambda}\bm{S}_{\textbf{n},\lambda}^*(\br)\right]\notag\\
\label{eq:vectorpot}
&\bm{S}_{\textbf{n},\lambda}(\br) = \bm{\epsilon}_{\mathbf{n},\lambda} e^{i\mathbf{k}_\mathbf{n}\cdot\mathbf{r}}/\sqrt{L^3}~.
\end{align}
Here we defined the transversal polarization vectors $\boldsymbol{\epsilon}_{\mathbf{n},\lambda}$ for mode and polarization $(\mathbf{n},\lambda)$~\cite{greiner1996}, and the creation and annihilation operators can be expressed in terms of displacement coordinates $q_{\mathbf{n},\lambda} = \sqrt{\frac{\hbar}{2\omega_{\mathbf{n}}}} (\hat{a}^\dagger_{\mathbf{n},\lambda}+\hat{a}_{\mathbf{n},\lambda})$ and their conjugate momenta $-i\partial_{q_{\mathbf{n},\lambda}} = i \sqrt{\frac{\hbar\omega_{\mathbf{n}}}{2}}(\hat{a}^\dagger_{\mathbf{n},\lambda}-\hat{a}_{\mathbf{n},\lambda})$ of harmonic oscillators with the allowed frequencies $\omega_{\mathbf{n}} = c|\mathbf{k}_{\mathbf{n}}|$.

The non-relativistic minimally coupled Hamiltonian (including $N_e$ electrons and $ N_n $ nuclei) in Coulomb gauge then reads with $\hat{j}^0(\textbf{r}) = c \hat{n}(\textbf{r}) = c \sum_{i=1}^{N_e+N_n} q_i \delta(\textbf{r}-\textbf{r}_i)$
\begin{align}
\hat{H} &= \sum_{i=1}^{N_e+N_n} \frac{1}{2m_i} \big( -i\hbar\mathbf{\nabla}_i - \frac{q_i}{c} \mathbf{\hat{A}}(\textbf{r}_i)\big)^2 + \hat{H}^{\perp}_{em} + \hat{H}_{int,\parallel} \notag\\
&\hat{H}^{\perp}_{em} = \frac{1}{2} \sum_{\mathbf{n},\lambda} \left[ (-\imagi \partial_{q_{\mathbf{n},\lambda}})^2 + \omega_{\mathbf{n}}^2 q_{\mathbf{n},\lambda}^2 \right] \label{eq:mincH}\\
&\hat{H}_{int,\parallel} = \frac{1}{8 \pi \varepsilon_0} \sum_{i\neq j}^{N_e+N_n} \frac{q_i q_j}{\vert \textbf{r}_i -\textbf{r}_j \vert}~.\notag
\end{align}
Each charged particle then evolves under the influence of the kinetic-energy operator $(-\imagi \hbar \mathbf{\nabla} - \tfrac{q}{c} \mathbf{\hat{A}}(\br) )^2$ and at the same time experiences the instantaneous longitudinal field (Coulomb potential $\hat{H}_{int,\parallel}$) created by all the other charged particles. 
The nonrelativistic limit of the minimal coupling procedure leads therefore naturally to the appearance of a quadratic term (see also App.~\ref{app:nlimit}). This quadratic term provides the diamagnetic shift~\cite{faisal1987} of the bare modes and introduces a lowest allowed frequency~\cite{rokaj2019} which then removes the infrared divergence~\cite{spohn2004}. It is therefore not a drawback to have this term~\cite{vukics2014}. In contrast, it affects for instance optical spectroscopy \cite{forbes2016identifying,forbes2018role}, is responsible for
diamagnetism~\cite{stefanucci2013} and hence implies very important physical processes, such as the famous Meissner effect. Recent theoretical~\cite{todorov2010,schafer2019modification,rokaj2019} and experimental~\cite{george2016} studies focused on the ultra-strongly coupled light-matter dynamics as well as the prediction of enhanced electron-phonon coupling~\cite{sentef2018,thomas2019exploringsuperconductivity,hagenmueller2019enhancement} highlight the non-negligible influence of the collective diamagnetic shift.

Let us next, for convenience assume linear polarization $\bm{\epsilon}_{\mathbf{n},\lambda} = \bm{\epsilon}_{\mathbf{n},\lambda}^*$ and to connect to the more common formulation, switch to atomic units, such that $\epsilon_0=1/(4 \pi),~c=1/\alpha_0$ and the elementary charge $e=1$. There is a well-known procedure to connect to a Hamiltonian corresponding to the macroscopic Maxwell's equation by employing the unitary Power-Zienau-Woolley (PZW) transformation (see also App.~\ref{app:pzw})

\begin{align*}
\hat{U} = \exp\left(- \imagi \alpha_0 \int \diff^3r \mathbf{\hat{P}}(\br) \cdot \mathbf{\hat{A}}(\br) \right) 
\end{align*} 

for all charged particles contributing to the polarization
\begin{align*}
\mathbf{\hat{P}}(\br) &= - \sum_{j=1}^{N_e} \br_j \int_0^{1} \delta^3( \br - s \br_j) \diff s  
\\
 &+ \sum_{j=1}^{N_n} Z_j \textbf{R}_j \int_0^{1} \delta^3( \br - s \textbf{R}_j) \diff s
\end{align*} 
where the $j$th nucleus has the effective positive charge $Z_j$.
This implies that all physical charges contribute to the bound current such that $\mathbf{j}^{f}_{} = 0,~\nabla \cdot \mathbf{D} = \varepsilon_0 \nabla \cdot \mathbf{E} + \nabla \cdot \mathbf{P} = 0 $ and therefore $ \mathbf{D} = \mathbf{D}_\perp $.
However, since the vector-potential operator is purely transversal it also only couples to the transversal part of the polarization operator, which can be expressed in terms of the transversal delta-distribution~\cite{greiner1996} or we use the mode expansion of the vector potential directly. To do so we first, for notational simplicity, introduce the abbreviation $\alpha \equiv (\mathbf{n},\lambda)$, then use that the vector-valued functions $\bm{S}_{\mathbf{n},\lambda}(\br)$ in \eqref{eq:vectorpot} form a basis for the transversal square-integrable vector fields (see App.~\ref{app:trbasis}), and find with $S_{\alpha}(\br) = \bm{\epsilon}_{\alpha}\cdot \bm{S}_{\alpha}(\br)$ 
\begin{align}
\label{eq:polarization}
\mathbf{\hat{P}}_{\perp}(\br) &=  - \sum_{j=1}^{N_e} \sum_{\alpha} \boldsymbol{\epsilon}_{\alpha} \left(\br_j \cdot \boldsymbol{\epsilon}_{\alpha} \right) \int_0^{1} S^*_{\alpha}(\br) S_{\alpha}(s\br_j) \diff s 
\\
&+ \sum_{j=1}^{N_n} Z_j \sum_{\alpha}\boldsymbol{\epsilon}_{\alpha} \left(\textbf{R}_j \cdot \boldsymbol{\epsilon}_{\alpha} \right) \int_0^{1} S^*_{\alpha}(\br) S_{\alpha}(s\textbf{R}_j) \diff s \nonumber.
\end{align} 
The resulting Hamiltonian~\cite{loudon1988,craig1998,salam2016non} has the advantage that it can be conveniently expanded in multipoles of the interaction. We note, however, that the validity of such an expansion depends critically on whether it is considered as a perturbation of the wavefunction or affecting the operator itself and subsequently its self-consistent solution (see App.~\ref{app:spectralfeatures}).
The mode-expansion provides a consistent regularization such that terms like $\mathbf{\hat{P}}_{\perp}(\br)^2$ are well-defined, a necessity when multiplying delta distributions (see App.~\ref{app:trbasis}). This avoids the usual auxiliary assumption that some of these terms, which contribute to the polarization self-energy, are only taken into account perturbatively~\cite{andrews2018perspective}. It furthermore highlights how the condition of transversality of the Maxwell field also affects matter-only operators like the polarization. 
So far, the only restriction we employed was that we considered non-relativistic particles. This simplification is, however, usually not yet enough to allow for practical calculations.
In the following we do not want to consider this more general case but assume only dipole interactions. This approximation is very accurate provided the dominating modes of the photon field have wavelengths that are large compared to the extend of the matter subsystem. In the multipole form of the non-relativistic QED Hamiltonian this means that we discard the integration over $s$ in our transformation and the polarization operator~\cite{loudon1988}. 
The Hamiltonian we then find is the same as the one that we get if we approximate $\mathbf{\hat{A}}(\br) \simeq \mathbf{\hat{A}}(\br_{\text{Matter}})$ for the bilinear and quadratic coupling terms. This does not restrict the form of the cavity modes itself but merely its spatial extension in relation to the matter subsystem. In practice where, e.g., an ensemble of molecules interacts with the cavity mode, this simplification can become questionable. Such an ensemble might extend over macroscopic scales such that individual molecules will experience different couplings. In the following we take $\br_{\text{Matter}}=0$ for simplicity such that $S_{\alpha}(0)$ is real and we can straightforwardly 
perform the unitary PZW, also referred to as length gauge, transformation
$\hat{H} = \hat{U} \hat{H} \hat{U}^\dagger$  with $ \hat{U} = e^{-i\sum_{\alpha}(\sqrt{4\pi}S_\alpha(0)\boldsymbol\epsilon_{\alpha} \cdot \big[- \sum_{j=1}^{N_{e}}\textbf{r}_{j}+\sum_{j=1}^{N_{n}}Z_j\textbf{R}_{j} \big] ) q_\alpha }$. We accompany this by a canonical transformation which swaps the photon coordinates and momenta $ -\imagi \partial_{q_\alpha} \rightarrow -\omega_\alpha p_\alpha, ~q_\alpha \rightarrow -\imagi\omega_\alpha^{-1} \partial_{p_\alpha} $ while preserving the commutation relations~\cite{tokatly2013}. The non-relativistic QED Hamiltonian then reads
\begin{align}\label{eq:hamiltonian}
\hat{H} &= \hat{H}_{n} + \hat{H}_{e} + \hat{H}_{ne} + \hat{H}_{p} + \hat{H}_{ep} + \hat{H}_{np},
\end{align}
where the nuclear Hamiltonian is
\begin{align*}
\hat{H}_{n} &= \hat{T}_n + \hat{W}_{nn} = -\sum_{j=1}^{N_n} \frac{1}{2 M_j}\nabla_{\textbf{R}_j}^2 + \frac{1}{2}\sum_{i \neq j}^{N_n} \frac{Z_{i}Z_{j}}{\vert \textbf{R}_i - \textbf{R}_j \vert}
\end{align*}
with the bare nuclear masses $M_j$. The electronic Hamiltonian is
\begin{align*}
\hat{H}_{e} &= \hat{T}_e +\hat{W}_{ee} = -\frac{1}{2m_e}\sum_{j=1}^{N_e} \nabla_{\textbf{r}_j}^2 + \frac{1}{2}\sum_{i \neq j}^{N_e} \frac{1}{\vert \textbf{r}_i - \textbf{r}_j \vert}
\end{align*}
with the bare electron mass $m_e$ and the nuclear-electron interaction is given by
\begin{align*}
\hat{H}_{ne} =  -\sum_{j=1}^{N_n}\sum_{i=1}^{N_e}\frac{Z_{j}}{\vert \textbf{r}_i - \textbf{R}_j \vert} ~.
\end{align*}
Further, the photonic contribution for $M_p$ modes is then given by
\begin{align}\label{eq:fullcoupling}
\hat{H}_{p} + \hat{H}_{ep} + \hat{H}_{np} = \frac{1}{2}\sum\limits_{\alpha=1}^{M_p}\left[ -\partial_{p_\alpha}^2 + \omega_{\alpha}^2 \left(p_{\alpha} -\frac{\boldsymbol\lambda_{\alpha}}{\omega_{\alpha}} \cdot \textbf{R} \right)^2 \right], 
\end{align}
which incorporates the total dipole $\textbf{R} = - \sum_{j=1}^{N_{e}}\textbf{r}_{j}+\sum_{j=1}^{N_{n}}Z_j\textbf{R}_{j}$
of electrons and nuclei. The resulting bilinear coupling or $\textbf{R}$ itself might be occasionally defined with the opposite sign as a consequence of the inversion symmetry of Eq.~\eqref{eq:hamiltonian}.   Even if we break the inversion symmetry of the matter Hamiltonian as e.g. in Sec.~\ref{translational}, the photonic symmetry $p_\alpha \leftrightarrow -p_\alpha$ will retain the trivial connection between both Hamiltonians and their respective observables.
Here $M_p$ is a finite but arbitrarily large amount of photon modes which are the most relevant modes but in principle run from the fundamental mode of our arbitrarily large but for simplicity finite quantization volume up to a maximum sensible frequency, for example, an ultra-violet cut-off at rest-mass energy of the electrons (an extended discussion can be found in App.~\ref{app:trbasis}). The operator Eq.~(\ref{eq:fullcoupling}) contains the bilinear matter-photon coupling and the quadratic dipole self-energy term $\frac{1}{2\varepsilon_0}\int d\textbf{r} \hat{\textbf{P}}_\perp^2(\br) \rightarrow \frac{1}{2} \sum_{\alpha=1}^{M_p}(\boldsymbol\lambda_\alpha\cdot \textbf{R})^2$.

The fundamental light-matter coupling to mode $\alpha$ is then denoted by
\begin{align}\label{eq:fundamentalcoupling}
\boldsymbol\lambda_\alpha = \sqrt{4\pi} S_\alpha(0) \boldsymbol{\epsilon}_{\alpha},
\end{align}
which depends on the form of the mode functions and the chosen reference point for our matter subsystem~\cite{ruggenthaler2014,pellegrini2015}. 
This can lead to an increase of the fundamental coupling to a specific mode and is an inherent feature of the physical set-up, e.g., the form and nature of the cavity. In the following we will treat $\bm{\lambda}_{\alpha}$ and the corresponding frequencies $\omega_\alpha$ as parameters that we can adopt freely to match different physical situations, motivated by the recent experimental progress to sub-wavelength effective cavity volumes~\cite{chikkaraddy2016,baranov2017novel}. This also highlights that the self-energy term depending on $\boldsymbol\lambda_\alpha$ is influenced directly by the properties of the cavity, i.e, obtains increasing relevance with decreasing effective mode volume $S_\alpha(0) = 1/\sqrt{L^3}$ and increasing number of participating modes $M_p$.

Importantly, since the PZW gauge is equivalent to the Maxwell's equation in matter as introduced earlier, we now work in terms of the, purely transversal, displacement field~\cite{craig1998,rokaj2017,andrews2018perspective}
\begin{align*}
\mathbf{\hat{D}}_\perp = \sum_{\alpha} \frac{\omega_{\alpha}}{4 \pi}  \boldsymbol{\lambda}_{\alpha} p_{\alpha} 
\end{align*} 
and the transversal polarization operator
\begin{align*}
\mathbf{\hat{P}}_{\perp} = \sum_{\alpha} \frac{1}{4 \pi} \boldsymbol{\lambda}_{\alpha} \left(\boldsymbol{\lambda}_{\alpha} \cdot \textbf{R} \right).
\end{align*}
By construction, the electric-field operator in PZW gauge, no longer representing the conjugate momentum, becomes
\begin{align}
\label{eq:Efield}
\mathbf{\hat{E}}_{\perp} = 4\pi \left(\mathbf{\hat{D}}_\perp - \mathbf{\hat{P}}_{\perp} \right).
\end{align}
The combination of PZW and canonical-momentum transformation changed our canonical operators $\mathbf{\hat{B}} \propto -i\partial_{p_\alpha},~ \mathbf{\hat{D}}_\perp \propto p_{\alpha}$ and consequentially the representation of our original creation and annihilation operators to
\begin{align}
\begin{split}\label{anihilationcreation}
\hat{a}_{\alpha}&=\frac{1}{\sqrt{2\omega_{\alpha}}}\left(-\imagi \partial_{p_\alpha}-\imagi\omega_{\alpha}p_{\alpha}+\imagi\boldsymbol{\lambda}_{\alpha}\cdot \mathbf{R}\right), \\
\hat{a}^{\dagger}_{\alpha}&=\frac{1}{\sqrt{2\omega_{\alpha}}}\left(-\imagi \partial_{p_\alpha}+\imagi\omega_{\alpha}p_{\alpha}-\imagi\boldsymbol{\lambda}_{\alpha}\cdot \mathbf{R}\right)~.
\end{split}
\end{align}	
We might yet again define a new harmonic oscillator algebra solely in terms of our new canonical operators $-i\partial_{p_\alpha} = i \sqrt{\frac{\omega_\alpha}{2}} (\hat{a}^\dagger_{\alpha,PZW} - \hat{a}_{\alpha,PZW}),~p_{\alpha} = \sqrt{\frac{1}{2\omega_\alpha}} (\hat{a}^\dagger_{\alpha,PZW} + \hat{a}_{\alpha,PZW}) $ to reach a potentially more familiar representations in terms of different $\hat{a}_{PZW},~\hat{a}^\dagger_{PZW}$. However, we have to consider then that the expression of physical observables in terms of creation and annihilation operators is not invariant under the PZW transformation, i.e., $\hat{a}_{PZW} \neq \hat{a}$. Special care has to be taken on how we interpret observables and design possible approximations as otherwise unphysical consequences arise as highlighted explicitly in Sec.~\ref{edfield_section}. We also see that in accordance to the Maxwell's equation in matter, by working with $\hat{\mathbf{D}}_\perp$ we implicitly take into account the back-reaction (polarization) of matter on the electromagnetic field. The physical field is found with the constitutive relation of Eq.~\eqref{eq:Efield}. Finally we note that the PZW transformation has removed the explicit diamagnetic contribution of the current and the physical current is now equivalent to the paramagnetic current. The diamagnetic term has, however, not vanished but is contained in the introduced phase of the coupled light-matter wave function~\cite{rokaj2017}.  

Let us clearly state a warning at this point. Unitary equivalence or gauge invariance, which implies that we obtain the same predictions in Coulomb and PZW gauge, is only fulfilled when the full Hilbert space (full basis-set) is considered. Any approximation in the molecular or photonic space will violate this equivalence and therefore result in deviating predictions \cite{craig1998,schafer2018ab,bernadrdis2018breakdown,DiStefano2019,li2020electromagnetic}. We remain with three different strategies. We can acknowledge this failure and focus on reasonable domains in which predictions remain in reasonably close agreement, e.g. focus on resonant interactions \cite{craig1998,schafer2018insights}. Alternatively, we adjust the PZW transformation to compensate the reduced space accordingly \cite{DiStefano2019}, or consider as much of the space as possible which leads us into the realm of first-principles cavity QED. This break-down of gauge invariance can have very fundamental consequences, as the long-standing debate of the (non-)existence of a Dicke superradiant phase shows~\cite{knight1978,viehmann2011superradiant,vukics2015}. Disregarding quadratic contributions ($\hat{\mathbf{A}}^2/\hat{\mathbf{P}}_\perp^2$) will consequentially merely allow to obtain perturbatively similar results.

Finally, there is one subtle question left. If we consider many photon modes they give rise to the radiative losses, that is, they constitute the photon bath of the matter subsystem into which the excited states can dissipate their energy~\cite{flick2019light}. Vacuum fluctuations give rise to effects like spontaneous emission, that is, turning the discrete eigenstates of the closed system described by a Sch\"odinger equation into resonances with finite line width~\cite{flick2019light}. Selecting furthermore only one or a very limited set of modes $\alpha$ will restrict retardation effects and can lead to unphysical superluminal transfer appearing in the (deep) ultra-strong coupling regime~\cite{munoz2018resolution}. In this work we are, however, not interested in lifetimes but in equilibrium states of the coupled light-matter system. In this case we can instead of keeping many modes subsume the vacuum photon bath by renormalizing the bare masses $m_e$ and $M_j$ of the charged particles, i.e., we use the usual physical masses such as $m_e=1$ in atomic units, and only keep a few important modes that are enhanced with respect to the free-space vacuum.

We have seen that already several approximations have to be employed to arrive at the above Hamiltonian which represents the usual starting point of most considerations in cavity QED and cavity-modified chemistry. Each approximation restricts its applicability but the basic physical constraints, i.e., gauge and coordinate-system (basis-set) independence, existence of a ground state, and radiation-less eigenstates, are as of yet conserved. It is now subject of the following sections to emphasize that ignoring the transversal self-polarization $\frac{1}{2} \sum_{\alpha=1}^{M_p}(\boldsymbol\lambda_\alpha\cdot \textbf{R})^2$ or diamagnetic $\frac{q^2}{c^2}\hat{\textbf{A}}^2(0)$ terms will inevitably break some of those fundamental constraints. This implies by no means that perturbative treatments that ignore these terms, either by restricting the Hilbert space of the matter subsystem or by perturbation theory on top of free matter observables, might not provide accurate predictions~\cite{bernadrdis2018breakdown,schafer2018insights,Stokes2019}. It shows however that care has to be taken when the quadratic terms are disregarded.

\section{Necessity and implications of quadratic couplings in the dipole approximation}
\label{Necessity}

\begin{figure}[h]
	\begin{center}
		\includegraphics[width=1\linewidth]{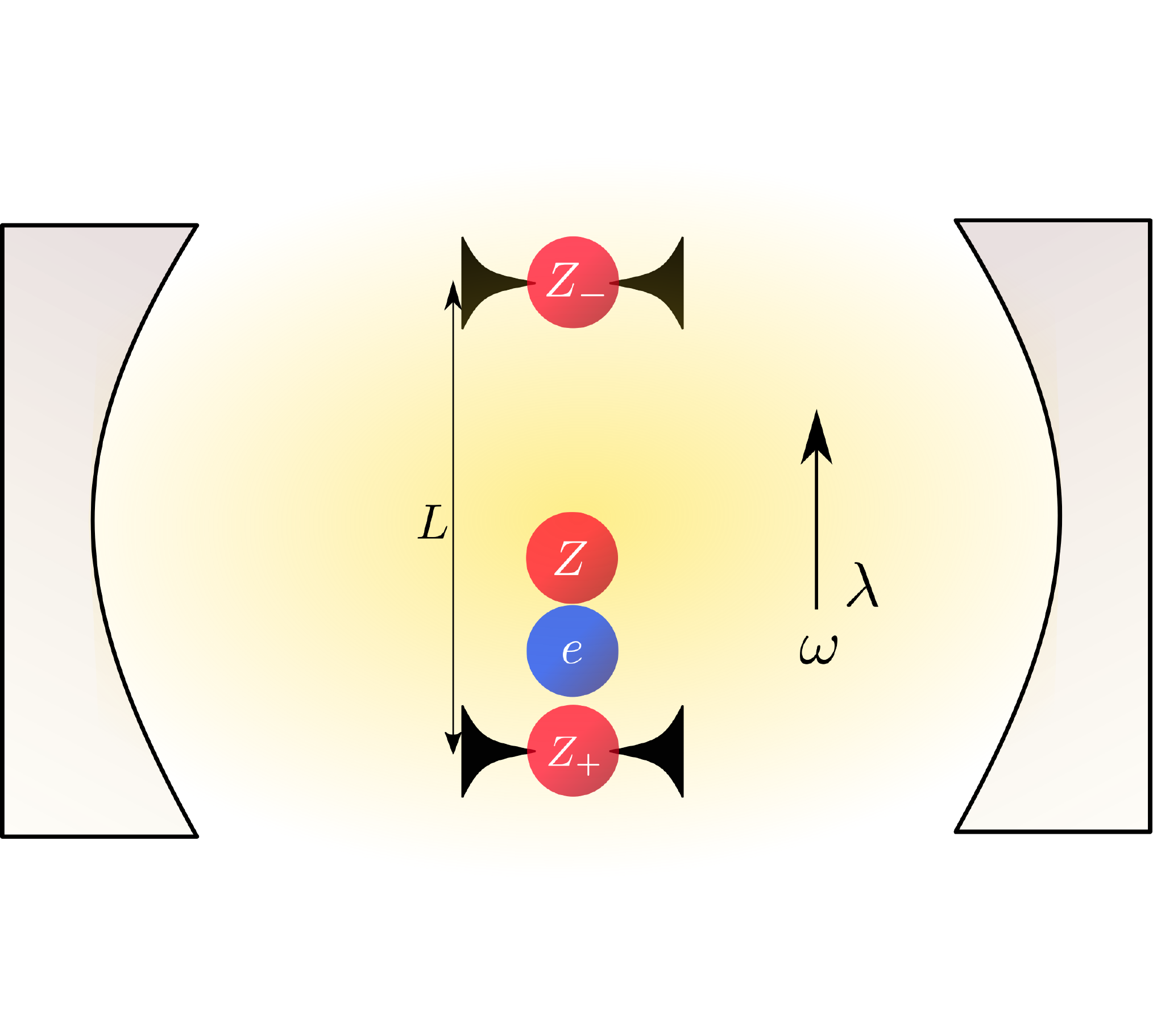}
		\caption{Illustration of the Shin-Metiu model insight the cavity. The two outer nuclei with the distance $L$ and charges $Z_{\pm}$ are fixed in position while central nucleus and electron can move freely within one dimension.} \label{shinmethiu}
	\end{center}
\end{figure}

Let us now consider concretely what happens if we discard the quadratic terms and which further physical constraints we violate. The example will be a simple molecular system, a slightly asymmetric one-dimensional Shin-Metiu model, coupled strongly to a single cavity mode as illustrated in Fig.~\ref{shinmethiu}. 
We subsume the rest of the photon bath in our description approximately into the physical mass of electron and nuclei. The Shin-Metiu model features one nucleus moving in between two pinned nuclei with in total one electron and is a paradigmatic model for non-adiabatic electron-nucleus coupling that gives rise to many interesting chemical processes. The Hamiltonian of this model with moving nuclear charge $Z=+1$ and removed vacuum shift $\frac{\omega}{2}$ is given by
\begin{align}
\label{eq:shinmetiu}
\hat{H} &= -\frac{1}{2M}\partial_X^2 -\frac{1}{2m_e}\partial_x^2 + V_{en_x}(x-X) \\
&+ Z_{-} V_{nn}(X-\frac{L}{2})+ Z_{+} V_{nn}(X+\frac{L}{2}) + Z_{-} V_{en}(x-\frac{L}{2}) \notag \\ \notag 
&+ Z_{+} V_{en}(x+\frac{L}{2}) + \frac{1}{2}\left[ -\partial_p^2 + \omega^2 \left(p -\frac{\lambda}{\omega} R \right)^2 - \omega\right], 
\end{align}
with the total dipole $R = -x + ZX$ and electron-nuclear and nuclear-nuclear potentials
\begin{align*}
V_{nn}(X \pm \frac{L}{2}) &= \frac{Z}{\vert X \pm \frac{L}{2}\vert}\\
V_{en}(x \pm \frac{L}{2}) &= \frac{erf\left[(\vert x \pm \frac{L}{2}\vert)/R_c\right]}{\vert x \pm \frac{L}{2}\vert}\\
V_{en_x}(x - X) &= \frac{Z~erf\left[(\vert x - X \vert)/R_f\right]}{\vert x - X\vert}
\end{align*}
where $erf$ represents the error-function.
For the following calculations we consider parameters $Z_{+}=1,~Z_{-}=1.05$, $M=1836~m_e,~L=18.8973,~ R_c=2.8346$,$~R_f=3.7795$~ with an electronic and nuclear spacing of $\Delta_x = 0.4,~\Delta_X=0.04$ between the equidistant grid-points and 40 photon number-states.
Furthermore, we couple electron and nucleus to a single cavity-mode with the frequency $\omega = 0.00231$, resonant to the vibrational excitation. We achieve ultra-strong vibrational coupling with $g/\omega=\lambda/\sqrt{2\omega} =0.40748$  in atomic units, where by no means the following results qualitatively depend on coupling or frequency. The strength of the light-matter interaction solely determines how quickly given effects will be visible and the selected values are close to those of previous publications in this field of research. It is important to realize that the associated wavelength to this frequency is $1.9724 \cdot 10^{5}~\text{\AA}~= 19.724~\mu m$ and thus differs by about four orders of magnitude from the computational box ($\approx 60~\text{\AA}$) that is considered for the matter system. Our example is thus safely within the validity of the long wavelength approximation when considering e.g. one-dimensional cavities.

\subsection{No bound eigenstates without self-polarization}\label{boundness}

\begin{figure}[h]
	\begin{center}
		\includegraphics[width=1\linewidth]{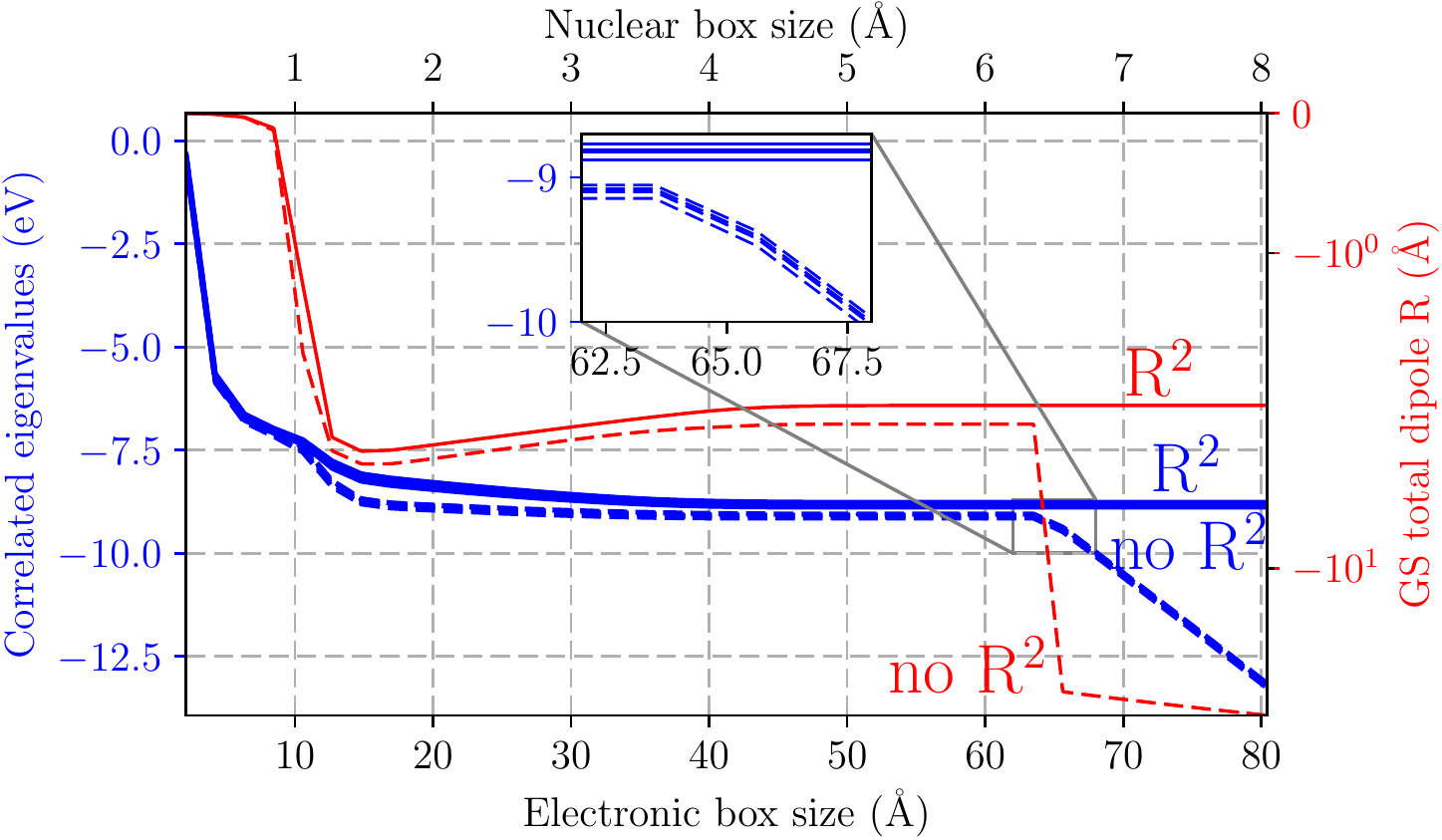}
		\caption{First four light-matter correlated eigenvalues with (blue, solid) and without (blue, dashed) self-polarization contribution as well as total dipole $R=-x+ZX$ with (red, solid) and without (red, dashed) self-polarization. While observables start to be converged with a box size around 50~\AA, without self-polarization the system starts to disintegrate already for slightly larger box-values as highlighted by the inset.} \label{boxsize_energy}
	\end{center}
\end{figure}

Let us start to investigate the most fundamental problem of discarding the quadratic term in the non-perturbative regime: The instability of the coupled system, i.e., that electrons and nuclei fly apart if coupled even in the slightest to the photon field unless we restrict the Hilbert space~\cite{rokaj2017,schafer2018insights}. To illustrate this we increase the simulation box stepwise by increasing the number of basis functions (grid-points), keeping other parameters fixed, and present first the light-matter correlated energies as well as the total dipole in Fig.~\ref{boxsize_energy}.
We find that by increasing the space of allowed wave functions, i.e., approaching the basis-set limit, the minimal-energy solution without the self-polarization term does not converge and minimizes the energy (dashed blue line) by increasing the total dipole (dashed red line). To put it differently, the system is torn apart and electrons occupy one side of the simulation box, while nuclei the other.

On the other hand, with the self-polarization term we see how we approach quickly the basis-set limit such that we have a basis-set independent result (red and blue solid line, respectively). The complete disintegration of the system without the self-polarization happens at a critical box size, which is just marginally larger than a box leading to converged results when the self-polarization is included. With further increasing box size, the energies resemble more and more those of an inverted shifted harmonic oscillator, which only supports scattering states~\cite{Chrucinski2003}. This illustrates that by a small ($\sim$20\%) variation of the simulation box we lose the physical character of the model and enter a non-physical regime. How drastic this effect will appear is given by the ratio of quadratically divergent potential energy $-\frac{1}{2}(\mathbf{\lambda}\cdot \mathbf{R})^2$ and the energy that is demanded to ionize the system from a given eigenstate $-\varepsilon_i$ (assuming non-interacting electrons for simplicity) such that a pure bilinear treatment would be perturbatively only reasonable for $-(\mathbf{\lambda}\cdot \mathbf{R})^2/2\varepsilon_i \ll 1$. In this sense the common ratio between coupling and excitation energy $g/\omega$, assuming resonance $\varepsilon_2-\varepsilon_1 = \omega$, can with slight adjustments to 
\begin{align}\label{eq:ec}
\text{Extension criterion} = \frac{\lambda^2}{4\varepsilon_i^2}~~~\text{(in atomic units)}
\end{align}	
be seen as an estimate how quick the given eigenstate "$i$" will become unstable without self-polarization component. This extension criterion \eqref{eq:ec} can be motivated by the single-particle Schr\"odinger equation $[\frac{-\hbar^2}{2m_e}\Delta + v(\textbf{r})]\Psi_i(\textbf{r}) =\varepsilon_i\Psi_i(\textbf{r})$ in the limit $r\rightarrow\infty,~v(\textbf{r})\rightarrow0,~\Delta\rightarrow\partial_r^2$ such that the long-range exponential decay of the state $\Psi_i \propto e^{-\sqrt{2m_e(-\varepsilon_i)}r/\hbar} = e^{-r/a_i}$ is defined by its characteristic extension $a_i = \hbar/\sqrt{2m_e(-\varepsilon_i)}$  (e.g. for the hydrogen atom the Bohr radius). The simulation box has to be large enough to at least fit the state "$i$" to an amount that we resolve an exponential decay $\sim e^{-1}$ (which is far from numerical convergence in fact). This provides an estimate of the extension of the eigenstate of interest and its associated self-polarization energy  $ (\mathbf{\lambda}\cdot \mathbf{R})^2 \approx (\lambda a_i)^2 = -\lambda^2/2\varepsilon_i $. While this might provide an orientation for theoretical calculations when instabilities are to be expected without self-polarization, even before the system is torn apart we see that the eigenvalues and the total dipole differ noticeably when increasing the basis set. Also other observables are changing without the self-polarization term, e.g., the non-perturbative Rabi splitting. The observables with self-polarization remain completely size-independent once reaching a sufficient basis-resolution.

\begin{figure}[h]
	\begin{center}
		\includegraphics[width=1\linewidth]{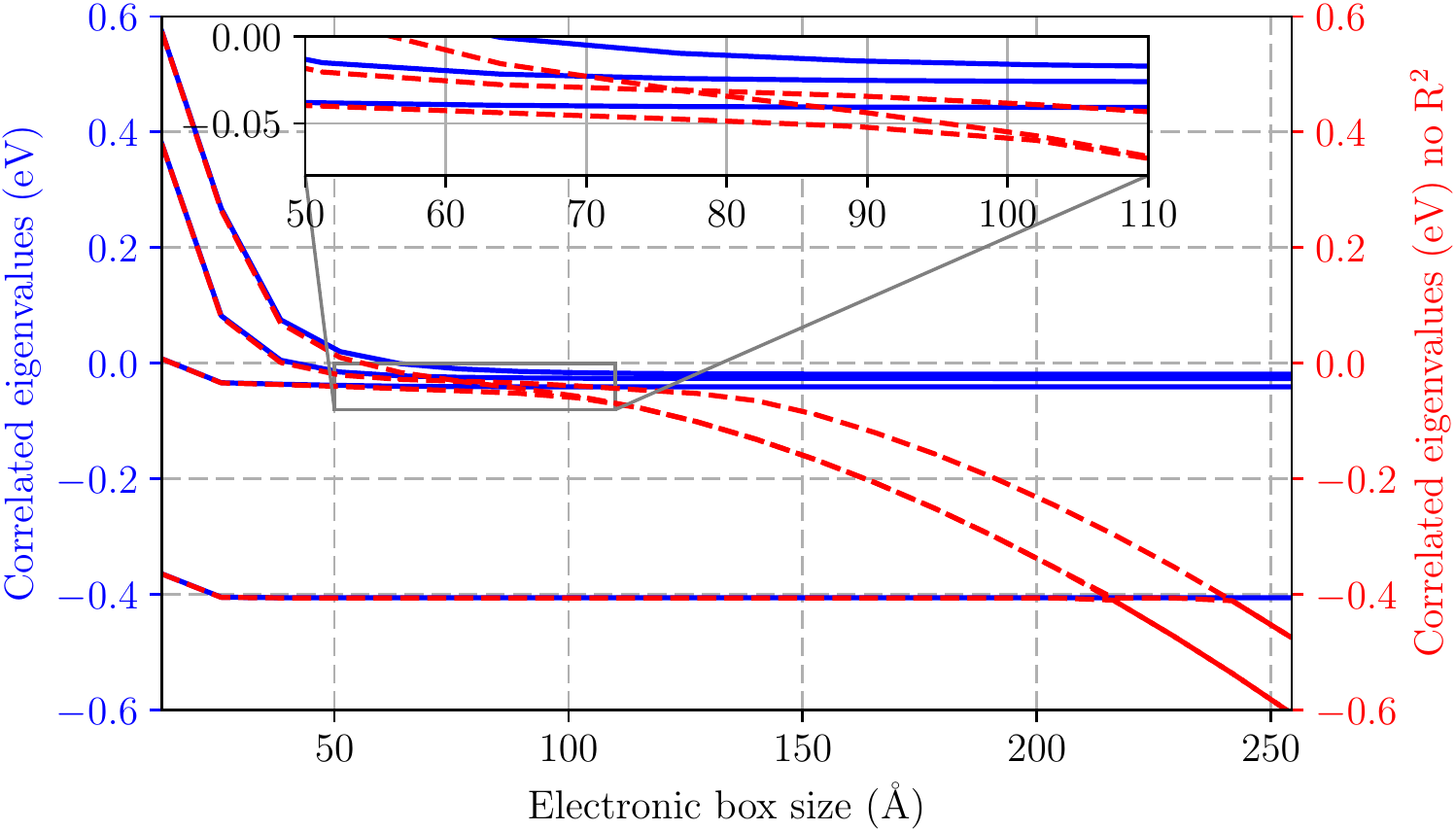}
		\caption{
			First four light-matter correlated eigenvalues with (blue, solid) and without (red, dashed) self-polarization contribution for the Rydberg-type weakly bound hydrogen model with grid spacing $\Delta x=0.8$ and 120 photon number-states. Until the box reaches large extends $\approx 200$ \AA~ the ground state without self-polarization is merely slightly deviating from the correct one. The excited states however, i.e., also the spectra and all observables involving excited states, are relatively weakly bound and experience unphysical behavior even before entering a converged regime. The inset magnifies the unphysical cross-over from physically bound into scattering states. The disintegration effect is qualitatively independent of the frequency.} \label{boxsize_energy_rydberg}
	\end{center}
\end{figure}

Let us illustrate how weakly bound states are affected with the help of a second numerical example. We select a simple one-dimensional soft-Coulomb Hydrogen atom but screen the nuclear charge $Z$ that binds the electron with $v(x) = -Z/\sqrt{x^2+1}$ to $Z = 1/20$. We couple this system rather weakly $g/\omega = 0.006$ with frequency $\omega = 0.01368$ in resonance to its first excitation (when converged) and as before increase step-wise the simulation box. Figure~\ref{boxsize_energy_rydberg} illustrates that although the ground state is merely perturbatively affected for up to 200 \AA, the excited states immediately turn into, for this case, unphysical scattering states. As before adding the self-polarization term will result in the expected spectrum, very much in contrast to the spectrum without the self-polarization component. The extension criterion $ \lambda^2/4\varepsilon_i^2$ leads for the ground state to $ 0.0011 $ and for the first excited state to $ 0.1664 $. While the ratio $g/\omega = 0.006$ gives the impression of rather weak coupling, the extension criterion provides a first indication that the excited states will be substantially affected by the self-polarization component.

While the bilinear interaction reduces the ground state energy with increasing coupling, the self-polarization contribution counteracts by an increase in energy and dominates for typical couplings the bilinear contribution, i.e., even the sign, thus the qualitative behavior, of energetic shift within the cavity can alter depending on the presence/absence of the self-polarization~\cite{schafer2018insights,flick2017}. This qualitative change is also represented in spatial observables, i.e., the self-polarization term favors a reduced polarizability and thus focuses charge density in domains where charge is already present~\cite{flick2017c,schafer2018insights,schafer2019modification}. The bilinear coupling, which furthermore scales with the frequency, is typically weaker affecting the ground state, features the contrary tendency and their competition will determine the qualitative distribution of charges inside the cavity~\cite{schafer2018insights,buchholz2019reduced}. The resulting consequences can e.g. include a reduced equilibrium bond length~\cite{flick2017,schafer2019modification} with an earlier onset of static correlation~\cite{schafer2019modification} that could be steered on demand by controlling the polarization of the field and therefore implies interesting opportunities for chemical considerations and electronic devices.

\begin{figure}[h]
	\begin{center}
		\includegraphics[width=1\linewidth]{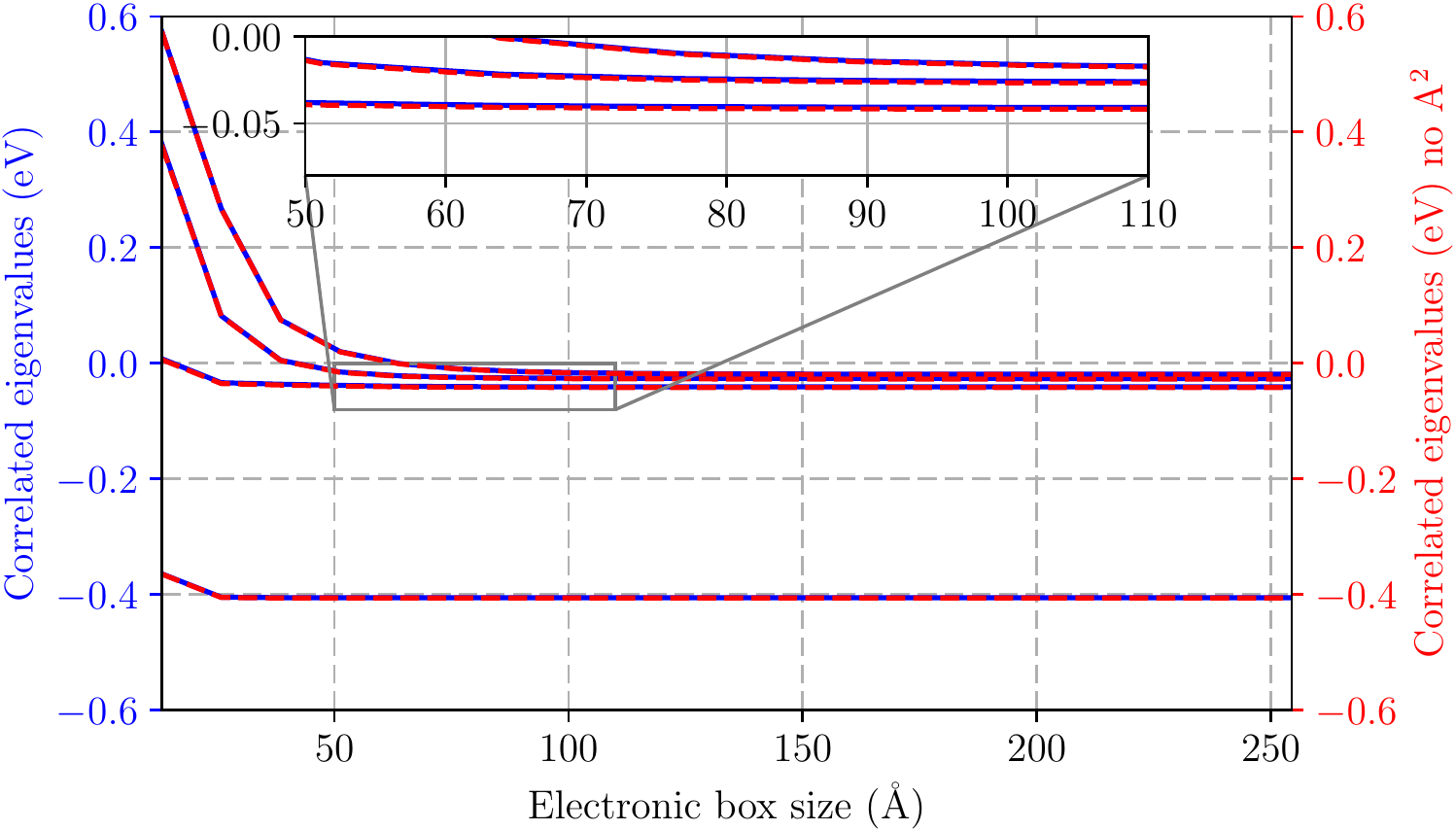}
		\caption{
			First four light-matter correlated eigenvalues in Coulomb gauge with (blue, solid) and without (red, dashed) diamagnetic $\hat{\textbf{A}}(\textbf{r})^2$ contribution for the Rydberg-type weakly bound hydrogen model (same parameters as Fig.~\ref{boxsize_energy_rydberg}). Disregarding the quadratic (diamagnetic) term omits this time the diamagnetic shift that leads to accurate agreement between both gauges. The inset magnifies the same region as in Fig.~\ref{boxsize_energy_rydberg}. Recall that the diamagnetic contribution attains increasing impact the smaller the frequency and the more polarizable matter is available.} \label{boxsize_energy_rydberg_coulomb}
	\end{center}
\end{figure}

Let us briefly inspect how the same system behaves in the Coulomb gauge instead. Fig.~\ref{boxsize_energy_rydberg_coulomb} illustrates the same correlated eigenvalues with increasing boxsize as the previous Fig.~\ref{boxsize_energy_rydberg}. Coulomb and PZW gauge lead to accurate agreement, a numerical difference in energy of less than $10^{-7}~eV$ for a box size of more than $40$ \AA, when both quadratic components are included. Omitting the diamagnetic term leads to a slight negative shift in the correlated eigenvalues and illustrates the relevance of the diamagnetic component, i.e., shifting the photonic excitations (see also App.~\ref{app:nlimit}). In the Coulomb gauge, couplings between higher excited states rescale lower matrix elements, demanding a well converged set of electronic eigenstates, and therefore the bilinear component accounts for the major effect of the self-polarization \cite{schafer2018insights,bernadrdis2018breakdown,li2020electromagnetic}. We should recall however, that the diamagnetic contribution scales with the amount of polarizable material and it attains increasing importance the smaller the frequency of the field. In this sense, performing the same investigation with a ten times smaller $\omega=0.00137=0.0372~eV$, the lowest excited states are photon replica with an energetic spacing between ground and first excited state of $0.0372~eV$ with diamagnetic contribution and $0.0254~eV$ without, highlighting a substantial deviation.

As a side remark, although the validity of the dipole approximation for high frequencies is questionable, the quadratic self-polarization term guarantees that the high-frequency photons essentially decouple from the matter subsystem. If only a purely linear-coupling is assumed, then the ultra-violet behavior is completely wrong as photons with arbitrarily high energies still interact with the matter subsystem~\cite{pellegrini2015}.

\subsection{Radiating eigenstates without self-polarization}\label{edfield_section}

\begin{figure}[h]
	\begin{center}
		\includegraphics[width=1\linewidth]{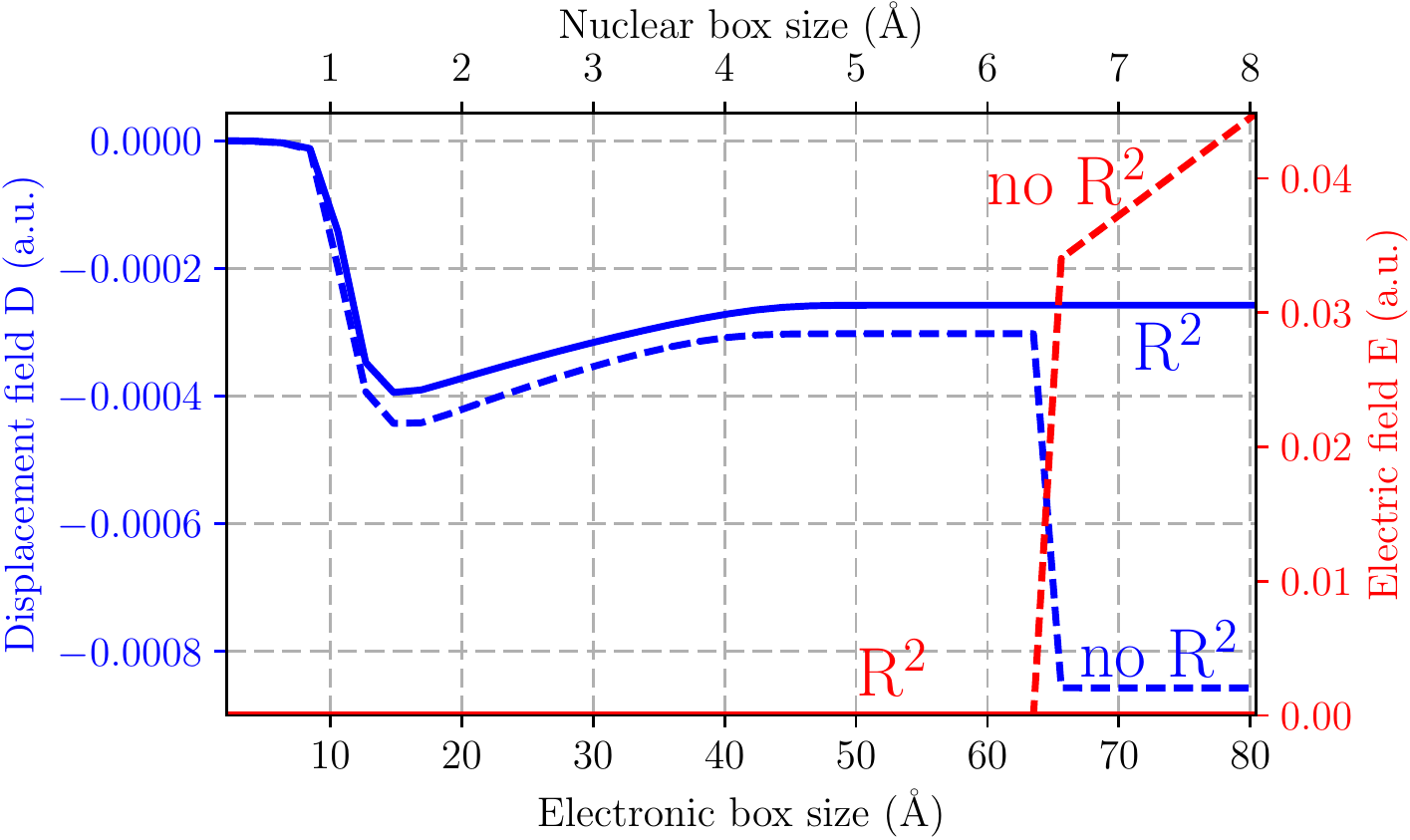}
		\caption{Displacement field with (blue, solid) and without (blue, dashed) self-polarization contribution as well as the electric field with (red, solid) and without (red, dashed) self-polarization. While the electric field is independent of the molecular convergence, therefore even in a restricted subspace we do not radiate and have a well defined equilibrium solution, the displacement field depends on the convergence of the molecular system.} \label{boxsize_edfield}
	\end{center}
\end{figure}

Let us look at another unphysical feature that appears when the self-polarization is neglected. In the case of the simple Rabi or Dicke model, where the particle is assumed perfectly localized (assuming effectively classical particles), the polarization is zero and we can associate the expectation values of the modes in PZW gauge with the electric field. However, if we consider an \textit{ab initio} treatment, this is no-longer the case and we need to use the correct definition of the electric field of Eq.~\eqref{eq:Efield}. In Fig.~\ref{boxsize_edfield} we then show the displacement as well as the electric field expectation values as we increase the number of basis functions.
This is again the same as allowing the electrons and nuclei to extend over an ever increasing spatial region which is equivalent to exploring the full Hilbert space. Also in these observables we see that the system with the self-polarization term included leads to simulation-box size independent results after roughly $\approx 45-50$ \AA~ (blue and red solid lines). Only extending the box slightly to $\approx 60$ \AA, the system without the self-polarization desintegrates (blue and red dashed lines). Recall here that this is well within the validity of the long-wavelength approximation as the matter and photon-field scales are separated by four orders of magnitude. By construction the system with the self-polarization term always obeys the basic constraint of zero electric field, while the system with only bilinear-coupling leads for large extensions to an eigenstate with non-zero electric field. This cannot be a physical ground state. 

\begin{figure}[h]
	\begin{center}
		\includegraphics[width=1\linewidth]{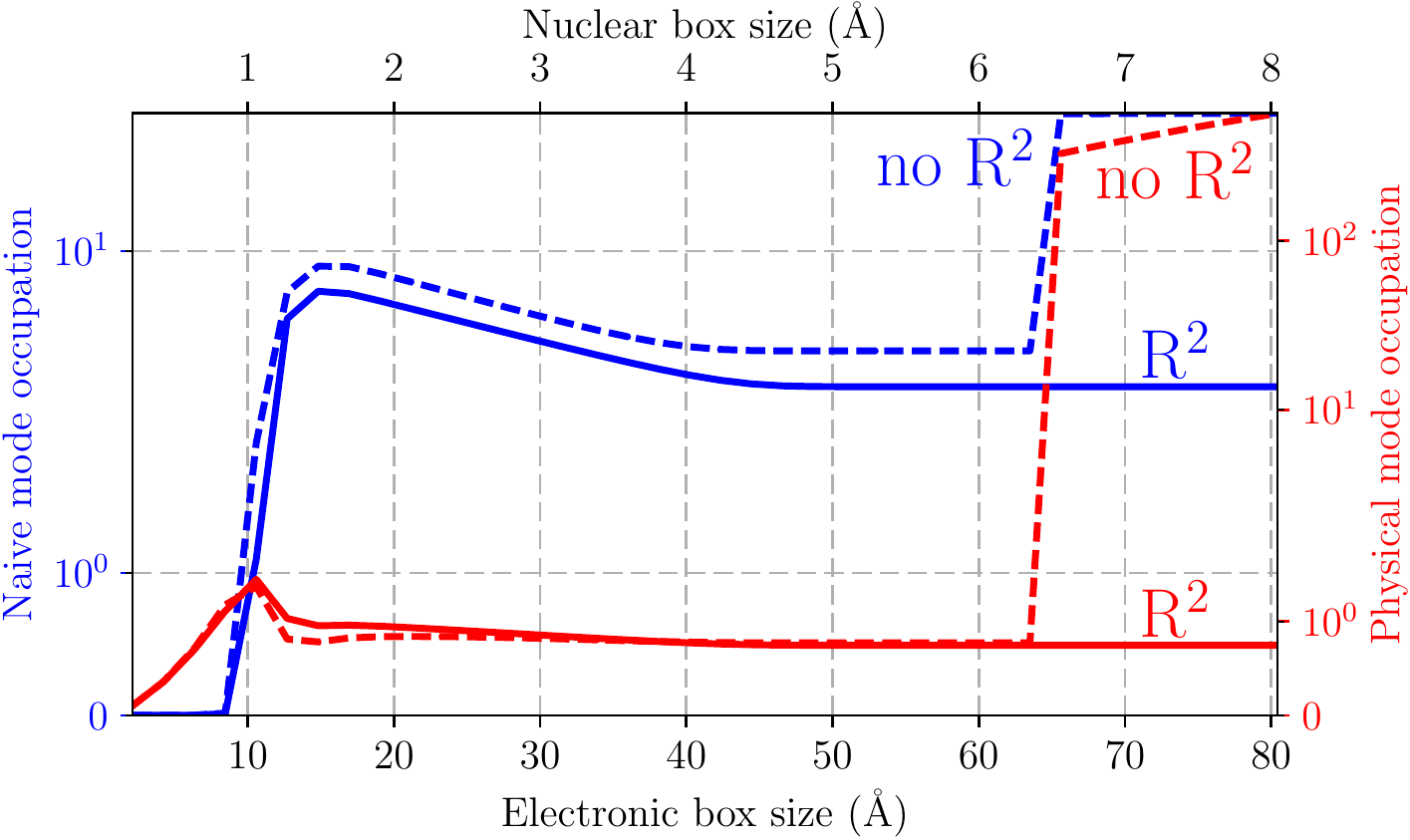}
		\caption{Naive $\hat{N}'$ (blue) and physical photon occupations $\hat{N}$ (red) in PZW gauge with (solid) and without (dashed) self-polarization contribution during the self-consistent solution.} \label{boxsize_phocc}
	\end{center}
\end{figure}

Realizing the connection between observable field and canonical momentum, let us turn our attention to the number of photons in the ground state. The photon number operator (the electromagnetic field occupation) is defined in Coulomb gauge as 
\begin{align*}
\hat{N} = \sum_{\alpha} \hat{a}^{\dagger}_{\alpha} \hat{a}_{\alpha}.
\end{align*}
In the PZW gauge these annihilation and creation operators are given by Eq.~(\ref{anihilationcreation}), and as a consequence the number operator $\hat{N}$ in this gauge is
\begin{align*}
	\hat{N}=\sum\limits_{\alpha=1}^{M_p}\left[ -\frac{1}{2\omega_{\alpha}}\partial_{p_\alpha}^2 + \frac{\omega_{\alpha}}{2} \left(p_{\alpha} -\frac{\boldsymbol\lambda_{\alpha}}{\omega_{\alpha}} \cdot \textbf{R} \right)^2 -\frac{1}{2} \right]. 
\end{align*} 
Originating from the change of conjugate momentum from electric to displacement field, we see that the self-polarization enters the definition of the photon number operator when we work in the PZW gauge. Without surprise, this leads to different occupations as if we would naively use
\begin{align*}
\hat{N}'=\sum_{\alpha=1}^{M_p}\left[ -\frac{1}{2\omega_{\alpha}}\partial_{p_\alpha}^2 + \frac{\omega_{\alpha}}{2} p_{\alpha}^2 -\frac{1}{2} \right]~,
\end{align*} and we illustrate this difference in Fig.~\ref{boxsize_phocc}. The alleged occupation $N'$ (blue) is higher than the physical occupation $N$ (red) caused by the permanent dipole.
Only for two-level models such as the Rabi model both definitions agree~\cite{schafer2018insights}.
Comparing Fig.~\ref{boxsize_edfield} and \ref{boxsize_phocc}, it is instructive to observe that displacement field $\textbf{D}_\perp$ and naive mode occupation $N'$ behave qualitatively very similar, obtaining relevant non-zero values only after a sufficiently large numerical box size is reached. In contrast the electric field $\textbf{E}$ remains system-size independent and the physical mode occupation $N$ adjusts merely quantitatively to the simulation box.
Not surprisingly, ignoring the self-interaction contributions in general leads to different results for different gauge choices.

\subsection{Coordinate system and dipole dependence without self-polarization}\label{translational}

The Hamiltonian of Eq.~\eqref{eq:hamiltonian} (and its variants) guarantees that all physical observables  in equilibrium are independent of the chosen coordinate system. 
This is obvious if we have a charge neutral system, where the Hamiltonian of Eq.~\eqref{eq:hamiltonian} is completely translationally invariant. This constraint is physically very reasonable, because without a spatial dependence, i.e., the manifestation of the long-wavelength approximation, the electromagnetic field cannot break the translational symmetry of the bare molecular system. If the system is not charge neutral, e.g., when we only consider electrons in an external binding potential, we do no longer have trivial translational invariance. To see this, consider a shift of the origin of the coordinate system along the polarization of the field such that also the total dipole operator $\textbf{R}$ is shifted. Note that this also changes the origin of the cavity as the long-wavelength approximation enforces that all molecules see the same field (of the now also shifted reference point) of the cavity. However, due to the zero-electric-field condition of a physical ground state we explicitly know the relation between the (shifted) dipole expectation value $\langle\textbf{R}\rangle$ of the matter subsystem and the expectation values of the displacement fields as $\langle p_\alpha \rangle = \frac{\boldsymbol\lambda_\alpha}{\omega_\alpha} \cdot  \langle\textbf{R}\rangle$~\cite{ruggenthaler2014,rokaj2017,schafer2018insights}. If we then further re-express the light-matter coupling with fluctuation quantities $\Delta \textbf{R} = \textbf{R} - \langle\textbf{R}\rangle$ such that Eq.~\eqref{eq:fullcoupling} becomes
\begin{align*} 
\frac{1}{2} \sum\limits_{\alpha=1}^{M_{p}} \left[-\partial_{p_\alpha}^2 + \omega_\alpha^2\left( \Delta p_\alpha - \frac{\boldsymbol\lambda_\alpha}{\omega_\alpha} \cdot \Delta\textbf{R}\right)^2\right],
\end{align*}
we find that in equilibrium the shifts cancel and the only remaining contribution in the Hamiltonian is given by the fluctuations around the mean values. Indeed, the equilibrium wave function in the new coordinate system is just the original wave function translated in space and the photon subsystem coherently shifted. 

As a consequence, the light-matter coupled system is invariant under shifts of the origin in equilibrium, no physical observable has a dependence on the permanent dipole.
That the equilibrium properties of light-matter coupled systems do not depend on a possible permanent dipole, 
is merely a consequence of how particles couple to the transversal electromagnetic field: only currents can interact with photons. A permanent dipole does only shift the photonic displacement field, which is not a physical observable, and the permanent dipoles of molecules will only contribute when the combined system is moved out of equilibrium.

\begin{figure}[h]
	\begin{center}
		\includegraphics[width=1\linewidth]{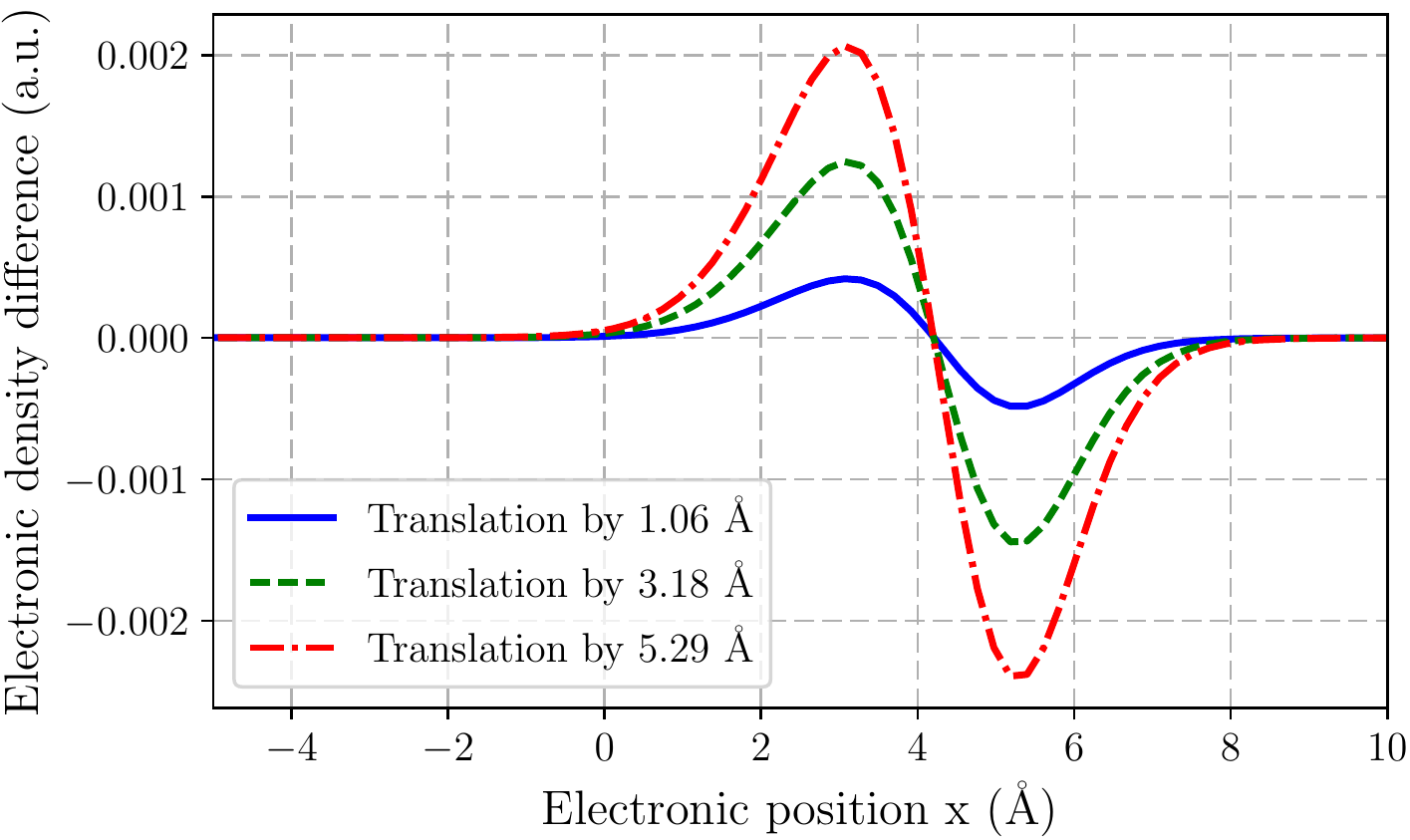}
		\caption{Electronic density difference between the translated (by a shift $\mu$) $n_{e}^{\mu}(x)$ and original system $n_e(x)$, i.e., $n^{\mu}_e(x-\mu) - n_e(x)$, without the self-polarization term. If the self-polarization is included the difference is always zero, i.e., the equilibrium-physics remains independent of the coordinate system and the permanent dipole. In the Shin-Metiu model the moving nucleus was slightly charged by $Z=+1.05$ and an electronic and nuclear box size of $59.27$ and $5.93$ \AA~ was chosen, i.e., before any scattering states appear. All other parameters remain as before.} \label{edens_shifted}
	\end{center}
\end{figure}

Only upon neglecting the self-polarization term can an unphysical dependence on the permanent dipole in equilibrium arise. To illustrate that even for small systems and shifts this can have a large influence, we consider the Shin-Metiu model from before, however, we slightly charge the complete system by using $Z=+1.05\vert e\vert$. We then perform a small shift $x \rightarrow x + \mu$ in the coordinate system, solve the corresponding Shin-Metiu model and determine the electronic ground state density $n_{e}^{\mu}(x)$ and then translate back, i.e., $n_{e}^{\mu}(x-\mu)$, to compare with the original (unshifted) solution $n_e(x)$.~\footnote{Note that in this case this is not a physical translation but merely a shift of origin for the coordinate system. However, within the long-wavelength approximation, physical (or \textit{active}) and coordinate system (or \textit{passive}) translations are synonymous and only when extending beyond this prototypical approximation observable differences between active and passive translations will appear.} As expected, when the self-polarization is included, we just recover the old density and $n^{\mu}_e(x-\mu) - n_e(x) \equiv 0$. In contrast, in Fig.~\ref{edens_shifted} we show the differences without the self-polarization and find an ever increasing difference for larger $\mu$ (increasing permanent dipole).  
The behavior of the Shin-Metiu model without the self-polarization is clearly unphysical, since observables should not depend on the coordinate system. Any approximate method tailored to perform self-consistent calculations should respect the above coordinate system independence by retaining the balance between bilinear and quadratic contributions. Consider for example the performance of the non-variational Krieger-Li-Iafrate approximation for \emph{ab initio} quantum electrodynamical density-functional theory presented in Ref.~\cite{flick2017c} which is breaking this balance. 

Notice that quadratic components also necessarily appear in other situations, i.e., they are indeed a quite general feature of non-relativistic Hamiltonians. If for example nuclear vibrations are approximated as phonon modes, the non-linear Debye-Waller-term $\propto \nabla_k \nabla_{k'} \hat{H} $ has to be added to the bilinear interaction~\cite{antonvcik1955theory,gonze2011theoretical}. This term originates from the quadratic elements in a Born-Huang expansion \cite{schafer2018insights} with the very same physical effects as the quadratic components $\hat{\textbf{A}}^2$ or $\hat{\textbf{P}}_{\perp}^2$, e.g., enforcing translational invariance and renormalizing the excitation energies.

\subsection{Collectivity, the limit of the Dicke model and plasmonic systems}

When considering a system of several molecules we can separate their instantaneous interactions mediated by longitudinal and transversal polarization fields (in the PZW gauge) by $\int d\textbf{r} \hat{\textbf{P}}^2 = \int d\textbf{r} \hat{\textbf{P}}_\parallel^2 + \hat{\textbf{P}}_\perp^2 $. Here the first term on the right-hand side corresponds to the Coulomb interaction, the second term corresponds to the self-polarization contribution~\cite{cohen1997photons}. We can further approximately distinguish between situations where the wavefunctions of the different constituents overlap strongly and situations where there is no strong overlap. The previous situation, often referred to as intra-molecular, demands to carefully consider Coulomb and self-polarization contributions simultaneously, where the Coulomb contribution is dominating in most situations. The latter situation of contact-free interactions, referred to as inter-molecular, between matter sub-systems leads to a perturbative (dipolar approximation) cancellation of instantaneous interactions such that  $\int d\textbf{r} \textbf{P}_{A}\cdot\textbf{P}_B\approx0$ and we are left approximately with purely bilinear and retarded interactions between those separated matter subsystems~\cite{power1982quantum}. This situation, however, does not allow to neglect longitudinal or transversal interactions when performing calculations locally for one of the sub-systems.
A consistent calculation considering, for example, molecular rearrangements during chemical reactions due to the influence of cavity-mediated strong light-matter coupling will thus demand also a consistent treatment of Coulomb and self-polarization contributions.

If we now enter into the realm where instantaneous contributions to the inter-molecular interaction cancel, it is often instructive to assume that indeed the local, i.e., sub-system, eigenstates are not affected by inter-molecular interactions and do not need to be updated during the process. In this case we can perform the pinned-dipole approximation which implies that each subsystem is localized at a specific position and distinguishable.
Starting from Eq.~\eqref{eq:hamiltonian}, we can then recover the Dicke model by absorbing the self-polarization contribution perturbatively by renormalizing the mass of the particles (similar to the perturbative treatment of the Lamb shift) such that the effective interaction reduces to the common bilinear coupling~\cite{craig1998}. 
In the case of the pinned-dipole approximation the bilinear coupling to the displacement field becomes equivalent to a coupling to the electric fields since the local polarization in $\textbf{E}_\perp = 4\pi(\textbf{D}_\perp - \textbf{P}_\perp) $ is zero by construction. To assume that the quantum sub-systems are perfectly localized and distinguishable is in stark contrast to a quantum-mechanical \emph{ab initio} description of molecules. Thus applying the Dicke model to deduce the influence of strong coupling on the local molecular states calls for a very careful analysis of all the applied approximations and their consequences. It furthermore permits physical features such as when charge-distributions start to overlap, as often the case in quantum chemical calculations, leading to a dependence of local observables on the surrounding (collective) ensemble~\cite{flick2017c}.

The occurrence of quadratic, i.e., Debye-Waller, terms in the electron-nuclei coupling highlights how general quadratic components are in a non-relativistic theory.
More closely connected to our current situation is the coupling to modes of a plasmonic environment. In principle, if we describe the plasmonic environment as part of the full system~\cite{jestadt2018real,Rossi2019}, the density oscillations of the plasmonic environment are captured in an \textit{ab initio} description by the Coulomb interaction and the induced transversal photon field and hence Eq.~\eqref{eq:hamiltonian} is directly applicable.
Let us assume, however, that we are not interested in a self-consistent calculation, can safely disregard contact-terms $\int d\textbf{r} \textbf{P}_{A}\cdot\textbf{P}_B=0$, and rather care about perturbative corrections. This consideration will lead to van-der-Waals (dipole-dipole) type interactions with different scalings in terms of their inter-molecular distance $R_{AB}$, independent of the choice of Coulomb or PZW gauge~\cite{craig1998}. The consideration of large distances subject to significant retardation are described by attractive Casimir-Polder interactions~\cite{casimir1948influence,salam2000contribution} scaling as $\mathcal{O}(R_{AB}^{-7})$. For smaller $R_{AB}$ retardation might be omitted and we enter the realm of instantaneous attractive interactions captured by the London dispersion potential $\mathcal{O}(R_{AB}^{-6})$. 
While those considerations are well tested and allow for excellent perturbative results, they would not allow a self-consistent calculation as these forces would merely result in a collapse of the wave function onto a singular point due to its unbalanced attraction. Assuming for instance a set of harmonic oscillators describing the plasmonic excitations coupled merely bilinear to the system of interest would introduce divergent forces $\propto -\sum_{\alpha}\mathbf{\lambda}_\alpha(\mathbf{\lambda}_\alpha\cdot \textbf{R})$ \cite{schafer2018insights}. Coulomb potential, wavefunction overlap and the Pauli principle give rise to repulsive components for small $R_{AB}$, modeled for example by the empiric $\mathcal{O}(R_{AB}^{-12})$ of the Lennard-Jones potential or the $\mathcal{O}(e^{-R_{AB}})$ of the Buckingham potential. It is therefore the higher-order components that are ensuring the stability of matter. A self-consistent treatment of molecules in a polaritonic cavity, which itself is modeled as, e.g., a simple harmonic oscillator~\cite{galego2019cavity,climent2019plasmonic}, thus needs to include higher-order couplings to describe a stable and physical system. 
Self-consistent calculations would therefore demand extending the quasistatic approximation \cite{englman1968optical,brako1975curvature,alpeggiani2014quantum} for plasmonic systems such that the plasmonic cluster responds to the coupled molecule. This is precisely the physical origin of the quadratic terms in QED, they allow the photonic or vice-versa the matter system to respond to the coupling by adjusting their excitation energies. For instance, the $\textbf{A}^2$ part can be subsumed into adjusted mode frequencies and further defines a minimal frequency, i.e., cures the infrared divergence, while the $\textbf{P}_\perp^2$ term renormalizes the energies of the material, all within the long-wavelength approximation. The very same effects should be present for a plasmonic cavity when consistently quantized. Such effects are already observed when performing \textit{ab initio} calculations with solely the longitudinal Coulomb interaction \cite{Rossi2019}. For small clusters, therefore small effective volume and high coupling strength, the modification of the response and volume due to the presence of the coupled molecule is non-negligible and modifies the plasmonic modes of the cluster. A purely bilinear coupling dictates entirely different physics (see App.~\ref{app:spectralfeatures} for a detailed discussion), violates all the aforementioned basic constraints and leaves such a simplification as inherently perturbative. While state-of-the-art models might provide insightful perturbative results, the development of corrected models should obtain additional interest and \textit{ab initio} calculations could prove beneficial to foster this effort.

\section{Summary}
\label{sec:summary}

It is the very nature of physics that our descriptions are necessarily approximate and that every theory has its limitations and drawbacks. And even if we have seemingly very accurate theories like QED, we need to reduce their complexity by employing further approximations and assumptions to render them practical. For QED this was historically done by employing perturbation theory and/or restrictions to a minimal set of dynamical variables. The resulting simplified versions of QED are due to their clarity and elegance a very good starting point for further investigations, provide for good reasons a common language for a variety of subjects and have provided tremendous insight over decades of research. Nevertheless, we need to be aware under which conditions these simplifications are valid and what their consequences are. With the recent experimental advances in combining quantum-optical, chemical and material science aspects~\cite{ebbesen2016} and the subsequent merging of \textit{ab initio} approaches with quantum-optical methods, it has become important to scrutinize these common assumptions~\cite{ruggenthaler2017b}.\\

In this work we have elucidated and illustrated the consequences of discarding quadratic terms that arise naturally in non-relativistic QED. Omitting them breaks gauge invariance, introduces a dependence on the coordinate system (or basis set), leads to radiating ground states, introduces an artificial dependence on the total dipole and in the basis-set limit leads to a disintegration of the complete system. However many of these effects can be mitigated if one works perturbatively or restricts the space. This is in accordance with many years of successful application of such approximations, but also highlights their limits of applicability. However, estimates of their applicability, such as the extension criterion Eq.~\eqref{eq:ec} discussed in Sec.~\ref{boundness}, become nowadays relevant for practical calculations. Certainly when strong coupling between light and matter modifies the local matter subsystem, as is suggested by recent experimental results~\cite{george2015,thomas2016,Lather2019,Thomas2019,Damari2019,thomas2019exploringsuperconductivity}, the quadratic terms can become important and determine the physical properties.\\

When looking beyond the simple Rabi splitting of spectral lines, which is the accepted indicator of the onset of strong coupling, other observables that contain further information about the matter subsystem should be able to highlight the necessity to modify the common Dicke or Rabi models, e.g., as demonstrated in Ref.~\cite{george2016}. By considering photonic as well as matter observables at the same time, the dipole-approximated bilinear coupling can be further scrutinized, the influence of quadratic coupling terms revealed and effects that are due to spatially inhomogeneous fields (beyond the long-wavelength approximation) observed. 
Furthermore, when the light-matter coupling renders bilinear, self-polarization and Coulomb interaction act on comparable energy scales, non-perturbative effects can be expected. At this point it is important to realize that this statement also holds spatially, i.e., that while a coupling might be considered small for certain bondlengths/extensions of the molecular system, at other parts or on other scales it might become substantial. The extension criterion Eq.~\eqref{eq:ec}, that weights the with spacial extension increasing divergent forces against the ionization energy, is motivated precisely with this spirit in mind, providing the to $g/\omega$ complementary parameter estimate. Consider e.g. the binding curve of a molecule, probing the dissociative regime with large distances will change the ratio of the aforementioned contributions until van-der-Waals type of interactions containing retardation effects have to be considered, a problem of also chemical interest \cite{andersson1998van}. Not just the equilibrium distance of molecules will change but especially their behavior in the stretched configuration will be effected \cite{flick2017,schafer2019modification}, a feature essential to describe chemical reactions. For relatively large systems, which are yet still small compared to the relevant wavelengths of the photon field, stronger effects would be expected. In the simple models we presented here we could have used a smaller coupling strength yet a spatially more-extended system and we would have found similar effects. Dynamics that probe the long-range part of potential-energy surfaces should be affected more strongly, especially true when compared to dynamics due to classical external laser fields in dipole approximation ignoring the self-polarization term. 
While our focus remained on the single molecular limit an additional essential scale of the system is represented by the amount of charged carriers, amplifying the dipole, polarizability, and therefore the self-polarization contribution. Extended systems (e.g. solids and liquids) and molecular ensembles with charge contact (e.g. biomolecules) are therefore expected to experience quite sizable influences by quadratic components, i.e., perturbatively seen renormalizing photonic or matter excitations due to the collective light-matter interaction~\cite{todorov2014dipolar,rokaj2017,schafer2018insights}. Recent investigations on cavity enhanced electron-phonon coupling~\cite{sentef2018,hagenmueller2019enhancement} and its role for superconductivity~\cite{thomas2019exploringsuperconductivity} might indicate the substantial scientific impact of this realization.
Exploring these situations where our theoretical descriptions begin to differ strongly thus hold promise of revealing further yet undiscovered effects~\cite{ruggenthaler2017b}.



\begin{acknowledgments}
	We would like to thank S. Buhmann, J. Feist, A. Salam and I. Tokatly for insightful discussions. This work was supported by the European Research Council (ERC-2015-AdG694097), the Cluster of Excellence 'Advanced Imaging of Matter' (AIM), Grupos Consolidados (IT1249-19), partially by the Federal Ministry of Education and Research Grant RouTe-13N14839, and the SFB925 "Light induced dynamics and control of correlated quantum systems".
\end{acknowledgments}



\appendix

\section{Fundamental coupling of light with matter and the emergence of diamagnetism}\label{app:nlimit}

Let us here briefly show how QED and its non-relativistic limit can be set up starting from classical electrodynamics. In vector-potential form
the microscopic description of the electromagnetic fields are given by
\begin{align*}
\mathbf{E}(\br,t) &= - \tfrac{1}{c} \partial_t \mathbf{A}(\br,t) - \mathbf{\nabla} A^{0}(\br,t),
\\
\mathbf{B}(\br,t) &=  \tfrac{1}{c} \mathbf{\nabla}\times \mathbf{A}(\br,t).
\end{align*}	
For later reference we use the vacuum permeability $\mu_0$ and vacuum permittivity $\epsilon_0$, which are connected to the speed of light by $c=1/\sqrt{\mu_0 \epsilon_0}$. If we then choose the, in the non-relativistic limit most convenient, Coulomb gauge
condition $\mathbf{\nabla}\cdot\mathbf{A}(\br,t) = 0$, the energy expression of the classical electromagnetic field is given by~\cite{greiner1996}
\begin{align}
\label{eq:MaxwellEnergy}
\mathcal{H}^{\perp}_{\rm{em}}(t) = \frac{\epsilon_{0}}{2} \int \mathrm{d}^3 r
\left( \mathbf{E}^2_{\perp}(\br,t) + c^2 \mathbf{B}^2(\br,t) \right) 
\end{align}
and the interaction among charged particles emerges via
\begin{align}
\label{eq:InteractionH}
\mathcal{H}_{\rm{int}}(t) &= \underbrace{- \frac{1}{c} \int \mathrm{d}^3 r \; \mathbf{j}(\br,t) \cdot \mathbf{A}(\br,t)}_{= \mathcal{H}_{\rm{int}, \perp}(t)}\\
&+ \underbrace{\frac{1}{2c} \int \mathrm{d}^3 r \; j^{0}(\br,t) A^{0}(\br,t)}_{= \mathcal{H}_{\rm{int}, \parallel}(t)} . \notag
\end{align}
Here we have used the decomposition of the electric field in a purely transversal part (polarized perpendicular to the propagation direction) $\mathbf{E}_{\perp}(\br,t) = - \frac{1}{c} \partial_t \mathbf{A}(\br,t)$ and a purely longitudinal part (polarized along the propagation direction) $\mathbf{E}_{\parallel}(\br,t) = - \mathbf{\nabla} A^{0}(\br,t)$. The electromagnetic field is coupled to a charge current $\mathbf{j}(\br,t)$ that obeys the continuity equation $\tfrac{1}{c} \partial_t j^{0}(\br,t) = -\mathbf{\nabla} \cdot \mathbf{j}(\br,t)$. We therefore see that it is the moving charges via their combined charge current that induce and modify the electromagnetic fields.

The above decomposition is furthermore very convenient to single out electrostatic contributions, which are given exclusively in terms of $\mathbf{E}_{\parallel}$ and $j^{0}$. With the Poisson equation in full space, 
which determines the zero component of the electromagnetic vector potential
\begin{align*}
A^{0}(\br,t) = \frac{1}{c}\int \rm{d}^3 r ' \frac{j^{0}(\br',t)}{4 \pi \epsilon_0|\br - \br'|},
\end{align*}
the term $\mathcal{H}_{\rm{int}, \parallel}(t)$ in Eq.~\eqref{eq:InteractionH} can be brought into the form
\begin{align}
\label{eq_Coulomb}
\mathcal{H}_{\rm{int}, \parallel}(t) = \frac{1}{2c^2} \int \int \mathrm{d}^3 r \; \mathrm{d}^3 r' \; \frac{j^{0}(\br',t) j^{0}(\br,t)}{4 \pi \epsilon_0|\br - \br'|}
\end{align}
and thus corresponds to the longitudinal Coulomb interaction, typically dominating the electronic  structure of condensed matter.
If we are not in vacuum on all of $\mathbb{R}^3$ but instead have, e.g., boundaries with certain boundary conditions, the Coulomb kernel $-\nabla^2 \tfrac{1}{4 \pi |\br - \br'|} = \delta^3(\br -\br')$ changes accordingly. This also changes the Coulomb interaction among charged particles, e.g., for cavity situations it can lead to the inclusion of mirror-charges, depending on the selected gauge \cite{power1982quantum}.
Further, due to the Coulomb-gauge condition, the first term on the right-hand side of Eq.~\eqref{eq:InteractionH} is merely coupling to the transversal part of the charge current and can thus be rewritten as
\begin{align}
\label{eq_Transverse}
\mathcal{H}_{\rm{int}, \perp}(t)  = -\frac{1}{c} \int \mathrm{d}^3 r \; \mathbf{j}_{\perp}(\br,t) \cdot \mathbf{A}(\br,t). 
\end{align} 
We have therefore divided the interaction due to coupling with a charge current into a purely longitudinal (electrostatic) and a purely transversal one. 
To quantize the theory we need to promote the classical vector potential to a quantum field $\hat{\mathbf{A}}(\br)$, which is basically a sum of quantum harmonic oscillators~\cite{greiner1996, ruggenthaler2014} (see also Eq.~\eqref{eq:vectorpot}).
The quantum fields include the transversal character via the transversal delta-function $\delta_\perp^{ij}$ in the commutation relations between conjugate fields. We furthermore
need to promote the classical charge current to the conserved charge-current operators $\hat{\mathbf{j}}(\br)$  and $\hat{j}^0(\br) = c \hat{n}(\br)$ of the non-interacting matter subsystem~\cite{greiner1996, ruggenthaler2014}. In this way the total charge current of the quantized particles generates the quantized electromagnetic field, and at the same time the photon field modifies the movement of the quantized particles. 
Hence QED becomes a self-consistent theory of light and matter, and equilibrium is reached when a force balance among the constituents is reached.\\
This clear procedure holds true if we consider QED with Dirac particles and thus the Dirac current. If we, however, take the non-relativistic limit for the particles and thus also for the conserved charge current, a subtlety arises with important consequences. By expressing the positronic degrees of freedom to first order in $1/m c^2$ in terms of the electronic components, a term quadratic in the vector potential appears~\cite{Engel2011, ruggenthaler2014}. This means that in Eq.~\eqref{eq_Transverse}, if we use the conserved current $\mathbf{j}(\br,t) = \mathbf{j}_{\rm p}(\br,t) + \mathbf{j}_{\rm d}(\br,t)$ that consists of the paramagnetic current $\mathbf{j}_{\rm p}$ plus the diamagnetic current $\mathbf{j}_{\rm d}$~\cite{stefanucci2013}, a correction term of the form $-\tfrac{1}{c}\int j^{0}(\br,t) \tfrac{q}{2 m c^2} \mathbf{A}^2(\br,t)$ has to be added~\cite{ruggenthaler2014}. This leads to the appearance of a \textit{quadratic coupling} term. This quadratic term renders the coupling defined by Eq.~\eqref{eq_Transverse} consistent with the minimal coupling prescription also in the non-relativistic limit (see the usual minimal-coupling form of Eq.~\eqref{eq:mincH}). Indeed, this extra term is due to the explicit appearance of the diamagnetic current contribution $\hat{\textbf{j}}_d (\br) =  -\frac{q}{m c}\hat{n}(\br) \hat{\textbf{A}}(\textbf{r})$~\cite{greiner1996, craig1998, ruggenthaler2014} that in the Dirac current arises only implicitly as can be seen by the Gordon decomposition~\cite{Engel2011, ruggenthaler2014}. This quadratic coupling term captures the effective photon-photon interaction due to the discarded positronic degrees of freedom. A direct beneficial consequence of this explicit diamagnetic term is that it removes the infrared divergence of relativistic QED~\cite{spohn2004}. This can be best understood by considering the Heisenberg equation of motion, analogue to the inhomogeneous microscopic Maxwell's equation,
\begin{align}\label{diamagneticshift}
\begin{split}
\left(\tfrac{1}{c^2}\partial_t^2 - \nabla^2\right) \hat{\textbf{A}}(\textbf{r},t) = \mu_0 c \hat{\textbf{j}}_\perp(\textbf{r},t)~.
\end{split}
\end{align}
Here $\hat{\textbf{j}}_\perp(\textbf{r})$ is the transversal part of the physical current operator $\hat{\textbf{j}}(\textbf{r}) = \hat{\textbf{j}}_{\rm p}(\textbf{r})+ \hat{\textbf{j}}_{\rm d}(\textbf{r})$. Grouping the diamagnetic current with the vector potential on the left-hand side shows that the mere existence of charged particles will modify the frequency of the bare fields (see also Sec.~\ref{Necessity} for the dipole case and recall Sec.~\ref{sec:Maxwell}).

Only when longitudinal \eqref{eq_Coulomb} and transversal \eqref{eq_Transverse} coupling are treated consistently they provide a local interaction. However, in practice often only one of the two interactions is treated explicitly depending on which properties of the combined light-matter system one is interested in~\cite{ruggenthaler2017b}. Focusing on quantum mechanics, e.g., the electronic structure as essential to describe chemical reactions, the transversal interaction is often omitted and one merely implicitly considers the fluctuations in form of the physical mass \cite{schafer2018insights,spohn2004}.
Quantum optical considerations on the other hand focus typically on the description of the transversal fields and thus strongly simplify the electronic structure. The resulting quantum-optical models are designed to predict specific photonic observables and are consequentially limited in their predictability for the matter subsystem~\cite{schafer2018insights,schafer2019modification}. Recent interest in strong-light matter interaction is calling now for a consistent treatment of those historically as complementary perceived limits.

Under certain assumptions the diamagnetic term can indeed be absorbed by a redefinition of the frequencies and polarizations of the field modes~\cite{faisal1987,todorov2014dipolar,schafer2018insights,rokaj2019}. These redefinitions depend on the matter subsystem (more specifically the number of charged particles) and lead to the diamagnetic shift of the photon field which can be observed experimentally~\cite{li2018, paravacini2019, todorov2010}.
Since the difference between the bare and the diamagnetically-dressed photonic quantities go as $\sqrt{N/V}$, where $N$ is the total number of particles and $V$ the quantization volume, it is often argued \cite{faisal1987} that one can use the bare quantities for finite systems. This is not entirely correct. The same argument would predict that the coupling between light and matter (see Eq.~\eqref{eq:fundamentalcoupling}) would be zero. The reason for non-zero coupling lies in the fact that by making the quantization volume larger (and approaching free space) the amount of modes in any frequency interval approaches infinity as well. So while indeed the coupling to an individual mode becomes zero, the coupling to the continuum of modes is non-zero. Thus when we keep individual modes in our theoretical description we effectively treat a small but finite frequency interval of modes. This frequency interval can be related to the effective mode volume, since it gives the spacing between the effective modes. Consequently we also need to dress the photon operators diamagnetically.

\section{The Power-Zienau-Wooley gauge transformation}\label{app:pzw}

While we here perform the unitary PZW transformation after having chosen the Coulomb-gauge quantization, which leaves the vector-potential operator invariant but leads to an adjusted conjugate photon-field momentum and coupling, one can equivalently use the PZW (multipolar) gauge of the field to perform the quantization procedure~\cite{babiker1983derivation}. This gauge is connected to the Coulomb gauge by adjusting the phase of each particle by $\theta(\textbf{r})=-q\alpha_0 \int_0^1 \textbf{r}\cdot\textbf{A}(s\textbf{r})ds$~\cite{babiker1983derivation}. This extra phase removes the explicit diamagnetic component from the physical current but also assigns a longitudinal component to the vector potential that can be associated with the Coulomb interaction. While the PZW gauge features, similarly to the Coulomb gauge, a purely transversal light-matter coupling, it mixes, in accordance with the macroscopic Maxwell's equations, light and matter degrees of freedom~\cite{power1959coulomb,woolley1980gauge,babiker1983derivation,cohen1997photons,andrews2018perspective}.

\section{Transversal basis and distributions}\label{app:trbasis}

For an arbitrary cavity geometry, the vector-valued eigenfunctions can be very complicated, deviating from simple plane waves. Still this basis $\bm{S}_{\alpha}(\textbf{r})$ is assumed to obey the condition $\nabla\cdot\bm{S}_{\alpha}(\textbf{r}) =0$. In this case we need to perform the mode expansion of the vector-potential operator and the polarization operator with the corresponding modes. Selecting a basis will define the representation of the delta distribution and the according polarization.
It is important to note that while there are many equivalent representations of the delta distribution, e.g., by using different basis sets $S_{\alpha}(\br)$, multiplications of distributions are not uniquely defined~\cite{schwartz1954limpossibilite,colombeau1985elementary}. Indeed, the origin of the divergence in quantum field theories stems from the fact that (operator-valued) distributions are multiplied~\cite{thirring2013quantum} and a regularization and renormalization procedure needs to be employed to give a finite answer. The usual way of regularization is equivalent to introducing a cutoff in the mode expansion $\alpha$ and hence by keeping this explicit instead of working with an unspecified representation of the delta distribution we avoid non-uniqueness problems~\cite{andrews2018perspective}.
We could straightaway also use the full infinite space~\cite{spohn2004}, but this will just make the notation unnecessarily complicated, as the above Hamiltonian converges in the norm resolvent sense to the infinite-space Hamiltonian for $M_p \rightarrow \infty$~\cite{arai1997existence}. Hence the above Hamiltonian can be made equivalent to the full infinite-space Hamiltonian if we increase the quantization volume.

\section{Spectral features of operators}\label{app:spectralfeatures}

Most arguments for performing a multipole expansion for a Hamiltonian are based on perturbation theory, where local properties derived from a fixed wave function do only slightly depend on higher-order contributions of a perturbing operator~\cite{craig1998}. This does, however, not mean that such arguments still apply for non-perturbative considerations, i.e., when the operator itself is changed and we solve the resulting equation self-consistently. Indeed, if we consider the influence of such expansions on the Hamiltonian directly, the opposite is usually true: The highest order determines all basic properties. An instructive example is a one-dimensional model atom with $\hat{H}= -\partial_x^2/2 + v(x)$ and $v(x)$ some binding potential centered at $x=0$ with $v(x)\rightarrow 0$ for $x\rightarrow \infty$. Its spectrum as a self-adjoint operator in $L^2(\mathbb{R})$ contains both, bound (eigenfunctions exponentially localized around $x=0$) and scattering (distributional eigenfunctions corresponding to the continuous spectrum) states. In such cases a harmonic approximation for certain ground-state properties where $v(x)\simeq v(0) + v'(0) x + v''(0) x^2/2$ is reasonable (assuming $v''(0)>0$) and the perturbative influence of higher-order terms proportional to $x^{n}$ with $n>2$ is minor. However, if we consider the actual Hamiltonian and treat higher-order terms proportional to $x^{n}$ non-perturbatively in $L^2(\mathbb{R})$, we see that either we have an operator that is unbounded from below (having no ground state) with purely continuous spectrum (no eigenfunctions but only scattering states) for $n$ odd \cite{rokaj2017}, or bounded from below with only bound states for $n$ even (again assuming that all $v^{(n)}(0)>0$). So all basic properties are only determined by the highest order of $n$. We therefore find that an expansion of an operator only becomes meaningful if we also indicate whether we consider it perturbatively for a fixed wave function or non-perturbatively. In this work we focus on the non-perturbative situation. Let us also mention that alternatively to perturbation theory a non-perturbative consideration but on a different Hilbert space of a restricted domain $\vartheta \subset \mathbb{R}^3$, i.e., we consider the operators on $L^2(\vartheta)$ with appropriate boundary conditions, or on a restricted state space $\{\Psi_1,...,\Psi_r\} \subset L^2(\mathbb{R})$ becomes possible (see also Sec.~\ref{Necessity}). As illustrated in Sec.~\ref{Necessity}, this will render the restricted domain or subset a relevant parameter for the theoretical prediction since the physical properties can then crucially depend on this parameter.

\bibliography{01_light_matter_coupling} 

\end{document}